\documentclass[apj]{emulateapj}

\usepackage{graphicx}

\newcommand{\kms}{km~s$^{-1}$}
\newcommand{\vsini}{$v\sin i$}
\newcommand{\vhelio}{$v_{\rm helio}$}
\newcommand{\msun}{$M_{\sun}$}
\newcommand{\rsun}{$R_{\sun}$}
\newcommand{\vmic}{$V-I_{\rm c}$}
\newcommand{\dupont}{du~Pont}
\shorttitle{Rotation of Open Cluster Giants}
\shortauthors{Carlberg}

\begin{document}

\title{Rotational and Radial Velocities of 1.3--2.2 \msun\  Red Giants in  Open Clusters}

\author{Joleen K. Carlberg\altaffilmark{1}}

\altaffiltext{1}{Department of Terrestrial Magnetism, Carnegie Institution of Washington, 
5241 Broad Branch Road, NW, Washington DC 20015, jcarlberg@dtm.ciw.edu}

\begin{abstract}
This study presents the rotational distribution of red giant stars (RGs)  in eleven old to intermediate age open  clusters.  
The masses of these stars are all above the Kraft break, so that they lose negligible amounts of their birth angular momentum (AM) during  the main sequence evolution.
However, they do span a mass range with quite different AM distributions imparted during formation, with the stars less massive than $\sim 1.6$\msun\ arriving on the main sequence with lower rotation rates than the more massive stars. 
 The majority of RGs in this study are slow rotators across the entire red giant branch regardless of mass,
 supporting the picture that intermediate mass stars rapidly spin down when they evolve off the main sequence and develop convection zones capable of driving a magnetic dynamo. Nevertheless, a small fraction of RGs in open clusters show some level of enhanced rotation, and  faster rotators are as common in these clusters as in the field red giant population. 
Most of these enhanced rotators appear to be red clump stars, which is also true of the underlying  stellar sample, while others are clearly RGs that are above or below the clump.
In addition to rotational velocities, the radial velocities and membership  probabilities of individual stars are also presented. 
Cluster heliocentric radial velocities for NGC 6005 and Pismis 18 are reported for the first time. 
\end{abstract}


\section{Introduction}
\label{sec:intro}

Studies of rotation and angular momentum (AM) evolution are becoming increasingly more sophisticated with the growth of  large observational datasets of stellar surface rotation probed by  both spectroscopic rotational velocities and photometric rotational periods.  Additionally, the exquisite photometric data from  \emph{CoRot} \citep{Auvergne:2009en}   and \emph{Kepler} \citep{2010Sci...327..977B} has enabled the first large scale studies of the interior rotation of stars other than the Sun.  Such internal studies of red giants (RGs) have clearly revealed differential rotation that changes over time.  \cite{Deheuvels:2012et} demonstrated that the cores of RGs rotate more rapidly than the surface, and  \cite{Mosser:2012dj} found that this core  rotation  slows significantly sometime during the late stages of the first ascent red giant branch (RGB) evolution. 
These new observations challenge our understanding of AM transport. 
\cite{2012A&A...544L...4E} and \cite{2012AN....333..971C} have shown that the physical mechanisms of transporting AM currently invoked in models (e.g., meridional circulation, shear mixing) are insufficient to explain the observed profiles.

Understanding the evolution of AM within a star throughout its life may help shine light on the unusually fast surface rotation that is sometimes found for apparently isolated RGs. In the context of RGs, ``fast'' rotation can refer to surface rotations as low as 4 or 5~\kms,  because the combination of AM shedding and a growing moment of inertia slows most RGs rotation to projected rotational velocities (\vsini) $\lesssim2$~\kms\ \citep{1996A&A...314..499D}.
The fast rotators are relatively rare, found only among a few percent of the field red giant population (e.g., \citealt{1993ApJ...403..708F}, \citealt{2008AJ....135..209M}, and \citealt{2011ApJ...732...39C}). The origin of this unusually high surface rotation is still a matter of debate, in part because  some relevant fundamental properties of field RGs are difficult to measure, especially masses. This is because  different stages of evolution for different masses can overlap in the Hertzsprung-Russell (HR) diagram both for  the same metallicity and over a range of metallicities.  These uncertainties are compounded by the fact that  errors in spectroscopically-determined parameters (such as [Fe/H]) grow as the stellar absorption lines become more rotationally broadened.
Knowing the stellar mass is important because of the very different rotation evolution followed by stars of different masses. Stars more massive than $\sim$1.3~\msun\ retain the majority of their birth AM  throughout their main sequence (MS) lifetimes, resulting in both large average rotational velocities and a large dispersion in rotational velocities (e.g., \citealt{Kraft:1967jh}, \citealt{Royer:2007dj}). 
Even within this intermediate mass regime, different rotational velocities are found. \cite{Gray:1982bc} inferred from the distribution of rotation with mass that the stellar  AM ($J$) followed a power law with mass ($M \propto J^{5/3}$), with a break to a steeper power law below 1.6~\msun. \cite{1997PASP..109..759W} demonstrated that the lower specific AM seen at 1.3--1.6~\msun\ was mostly imparted during the pre-MS stages, but also noted that some slow loss of AM might occur during the MS. 
Conversely, stars less massive than $\sim1.3$~\msun\  rapidly spin down, eventually erasing all information on their birth AM and  allowing the field of gyrochronology to use measured rotation as a proxy for age on the MS (e.g., \citealt{Barnes:2003ga}, \citealt{Barnes:2007fz}, \citealt{Chaname:2012jz}).

Given the larger AM seen in more massive MS stars, it seems logical to assume that the field population of fast RG stars are simply the most massive stars. One way of testing this assumption would be to measure the rotation of subgiant stars crossing the Hertzsprung gap, which would provide a direct measure of the spin-down from the MS to the base of the RGB.  However,  these stars are relatively rare.  To study a population of RGs with well-constrained masses, open cluster stars are needed.
To date, little attention has been paid to the distribution of \vsini\ of RGs in open clusters.  Therefore, in this study over 400 red giant candidate members of eleven open clusters  were chosen to probe the rotation
distribution in open clusters (Section \ref{sec:samples}). High resolution spectra were obtained for these candidates (Section \ref{sec:obs}), from which  heliocentric radial velocities (\vhelio) and \vsini\ are measured  (Section \ref{sec:measure}).
The \vhelio\ distributions are used to define the cluster's \vhelio\ and identify likely members (Section \ref{sec:vhelio}). The rotation distributions are analyzed with respect to both stellar masses and current evolutionary stages  (Section \ref{sec:vrot_results}), and potential explanations for the fastest rotators are explored (Section \ref{sec:sources}). Finally, the main results and conclusions of the study are summarized (Section \ref{sec:summary}).

\section{Cluster and Red Giant Selection}
\label{sec:samples}

\begin{deluxetable*}{lrrrrrr}
\tablecolumns{7}
\tablewidth{\textwidth}
\tabletypesize{\scriptsize}
\tablecaption{Literature Properties of Studied Open Clusters \label{tab:oc_list2}}

\tablehead{
   \colhead{Cluster Name} &
   \colhead{Age} &
   \colhead{$m-M$\tablenotemark{a}} &
   \colhead{[Fe/H]} &
   \colhead{E($B-V$)} &
   \colhead{Reference}&
   \colhead{ $M_{\rm RGB}$} \\
    \colhead{} &
   \colhead{(Gyr)} &
    \colhead{(mag)} &
   \colhead{(dex)} &
   \colhead{  } &
   \colhead{  } &
   \colhead{(\msun)}}
 
\startdata
Collinder 110		 & 1.41\tablenotemark{c}& 13.00\phm{$^c$}  &   \nodata   & 0.5\phm{$^c$}  & (1) &  1.9\tablenotemark{c} \\
   				 & \nodata & \nodata  &    $+0.03$\tablenotemark{c}  & \nodata & (5) & 1.9\phm{$^c$} \\  
				 & 1.1--1.5\phm{$^c$} & (11.8\tablenotemark{c} to 11.9)$_0$ &  0.00\tablenotemark{b} & 0.4\tablenotemark{c} & (10) & \nodata \\ 
Melotte 66			& 2.79\tablenotemark{c}& 13.62\phm{$^c$} &	 $-0.35$\tablenotemark{c}  &  0.143 & (1) & 1.4\tablenotemark{c}   \\ 
				& 5.33\phm{$^c$}& \nodata &	 $-0.38$\phm{$^c$}  &  \nodata & (2) & 1.2\phm{$^c$}  \\   
M67                	         &2.56\phm{$^c$} & 9.97\phm{$^c$}    & \nodata    &  0.059 & (1) &1.5\phm{$^c$} \\
                      		& 4.3\tablenotemark{c}  & (9.60\tablenotemark{c}$\pm$0.09)$_0$   &    $0.02$\tablenotemark{c}   & 0.04\tablenotemark{c}  & (2) & 1.3\tablenotemark{c} \\  
NGC 2477			&0.70\phm{$^c$} & 11.30\phm{c} & 	 $+0.01$\phm{$^c$} & 0.279\phm{$^c$} & (1) &  2.4\phm{$^c$} \\ 
				&1.00\tablenotemark{$^c$} & 11.45\tablenotemark{c}$\pm$0.08\phm{$^c$}  & 	 $0.00$\tablenotemark{$^c$} & 0.23\tablenotemark{c}  & (2) & 2.1\tablenotemark{c}   \\  
				& 1.00\phm{$^c$} &(10.5)$_0$ & 0.00\phm{$^c$} & 0.22--0.3\phm{$^c$} & (7) & 2.1\phm{$^c$} \\
NGC 2506			& 1.11\phm{c}& 12.95\phm{$^c$}  &	 $-0.37$\phm{$^c$} & 0.081 & (1) & 1.9\phm{$^c$} \\ 
 				& 1.99\tablenotemark{$^c$}& 12.65\tablenotemark{c} & Z=0.008\tablenotemark{$^c$} & 0.04\tablenotemark{c}  & (3)  & 1.6\tablenotemark{c}  \\ 
				& 2.14\phm{$^c$}& \nodata &	 $-0.42$\phm{$^c$} & \nodata & (2) & 1.5\phm{$^c$}  \\  
NGC 2660                  	& 1.08\tablenotemark{c}& 13.23\phm{$^c$} &    $-0.18$\tablenotemark{c} &  0.313 & (1) & 2.0\tablenotemark{c}    \\ 
                              	& 0.73\phm{$^c$}& \nodata &    $-0.55$\phm{$^c$} &  \nodata & (2)&  2.2\phm{$^c$} \\  
NGC 6005		 	& 1.20 & 13.54\phm{$^c$} & 0.00\tablenotemark{b}  &0.45\phm{$^c$} & (1) &  1.9\phm{$^c$}\\  
NGC 6134			& 0.93\tablenotemark{c}& 11.03\tablenotemark{c}&	  $+0.18$\tablenotemark{c} & 0.395\tablenotemark{c} & (1) & 2.2\tablenotemark{c}   \\  
					& 0.82--0.95\phm{$^c$}                  & (10.5)$_0$ & $+0.15$\tablenotemark{b}  & 0.35\phm{$^c$} & (8) & $\sim$2.2\phm{$^c$}\\
NGC 6253			& 5.01\phm{$^c$} & 11.51\phm{$^c$}&	 $+0.36$\tablenotemark{d} & 0.20\phm{$^c$} & (1) & 1.3\phm{$^c$}\\  
                                        & 3.5\tablenotemark{c}         & (11.15\tablenotemark{c})$_0$  &  Z=0.03\phm{$^c$}  & 0.25\tablenotemark{c}   &  (9) & 1.4\tablenotemark{c} \\
Pismis 18			&1.20\tablenotemark{c} & 13.3\phm{0$^c$}   & \nodata  &  0.50 & (1)         &  1.9\tablenotemark{c}  \\
				& \nodata & \nodata     &$0.00$\tablenotemark{c}   &  \nodata & (6)    & 1.9\phm{$^c$} \\
Ruprecht 147	           & 2.45\tablenotemark{c}  & 6.68\phm{$^c$}    &    \nodata & 0.15\tablenotemark{c} & (1) & 1.6\phm{$^c$}\\
		                   &  2.5\phm{0$^c$}&   7.35\tablenotemark{c}  &    $+0.08$\tablenotemark{c} & 0.25\phm{$^c$}  & (4)  & 1.5\tablenotemark{c}  
\enddata
\tablenotetext{a}{Values enclosed in parenthesis denote that the distance modulus has been corrected for extinction.}
\tablenotetext{b}{Adopted but not derived in the literature study.}
\tablenotetext{c}{Adopted in this paper.}
\tablenotetext{d}{This value is adopted for the cluster, but all isochrone-derived values are based on $Z=0.03$ ([Fe/H]$\sim+0.18$), the most metal-rich isochrone available.}

\tablerefs{(1)  WEBDA, (2) \citealt{Salaris:2004gh}, (3)  \citealt{Mermilliod:2007kv}, (4) \citealt{2013AJ....145..134C}, (5) \citealt{2010A&A...511A..56P}, (6) \citealt{Piatti:1998gs}, 
 (7) \cite{Eigenbrod:2004dx}, (8) \citealt{Ahumada:2013cp}, (9) \citealt{montalto09}, (10) \citealt{2003MNRAS.343..306B}  }
\end{deluxetable*}

The open cluster sample comprises the oldest, most nearby open clusters accessible in the Southern Hemisphere with the largest populations of RGs.  We compiled the sample by  querying the WEBDA database\footnote{Available online at: \url{http://www.univie.ac.at/webda/}.}  with the following criteria, (1) $\delta < +15^{\circ}$, (2) age $>$0.7 Gyr,  (3) distance modulus $< 14$, and (4) requiring either a list of red giant candidates or  color-magnitude diagrams (CMDs) for which the RGB  or red clump (RC) was clearly present.  The age constraint selects red giant masses less than $\sim 2.2$~\msun. 

\begin{figure}
\centering
\includegraphics[width=0.5\textwidth]{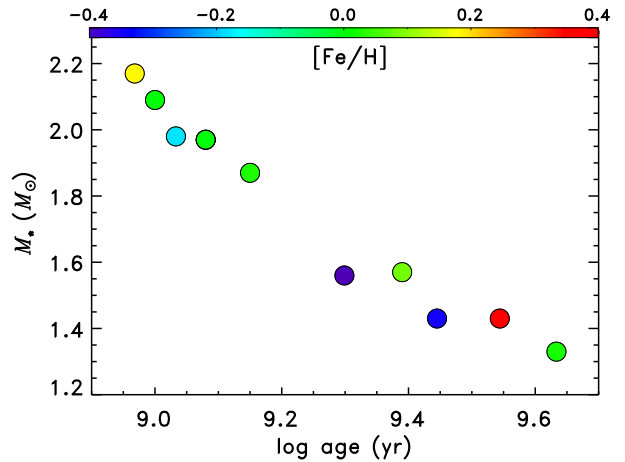} 
\caption{\label{fig:feh_age} Stellar ages and estimated RG masses of the open clusters in this study. The color scale gives the [Fe/H].  Masses come from \cite{Marigo:2008fy} isochrones.  NGC 6005 and Pismis 18 share a point in this plot at $\log$~age~$=9.07$, [Fe/H]=0,  and $M_\star=1.97$~\msun. (A color version of this figure is available in the online journal.)} 
\end{figure}

Table \ref{tab:oc_list2} lists the eleven clusters selected for this study, along with the clusters' age, distance modulus, metallicity, reddening, literature source, and estimated RG mass. Each cluster has at least one entry showing
the cluster properties available in the WEBDA database. Additional entries are given for clusters with multiple measurements in the literature. 
To maintain consistency with the original sources, the reported distance moduli  that have already been corrected for extinction effects are enclosed in ``()$_0$'' to distinguish them from distance moduli without such corrections.
The RG mass in the last column comes from the  \cite{Marigo:2008fy} isochrones matching the cluster properties. It refers to  the initial mass of the RG in the isochrone  that has just begun the core Helium burning stage 
(labeled ``BHeb'' in the isochrones). 
Cluster metallicities are generally given as [Fe/H], while the isochrones use metal mass fractions, $Z$. Conversions between the two come from Equations 10--11 in \cite{Bertelli:1994tq}. 
Some clusters have a significant variation in their reported properties. For example,  \cite{Salaris:2004gh} reports that NGC 2660 is much younger and more metal-poor than WEBDA. The choice of adopted cluster properties for this work are detailed in the individual cluster sections below.
The age, metallicity, and estimated RG mass for the eleven clusters in our study are displayed in Figure~\ref{fig:feh_age}.  NGC 6005 has no metallicity reported in the literature, and we adopt solar metallicity for convenience. As such, its age, metallicity, and RG mass are identical to that of Pismis 18. 

For each individual cluster, we compiled photometric data, positions, membership probabilities, etc.\ from WEBDA.   
 The amount of available data varied for each cluster; therefore, we address them individually below.  In general, we tried to select red giant stars photometrically and from the lists of red giant candidates in the cluster. Top priority  for observing was given to stars that appeared both in the WEBDA RG list and in our photometric giant list. We vetted these lists by looking at cluster membership probabilities, previously known \vsini,  and comparing radial velocities (RVs) to the accepted cluster velocities. We also eliminated spectroscopic binaries with periods shorter than $\sim$20 days, whose rotations may have been affected by tidal interactions. 
For three of the more nearby clusters, we also targeted some of the subgiant (SG) stars to probe the rotation rate before the stars reach the base of the red giant branch.

\begin{deluxetable*}{lrllrrrr}
\tablecolumns{5}
\tablewidth{0pc}
\tabletypesize{\scriptsize}
\tablecaption{Observed Red Giant Candidates \label{tab:oc_starlist}}
\tablehead{
   \colhead{Cluster } &
   \colhead{Star Number} &
   \colhead{RA} &
   \colhead{DEC} &
   \colhead{$V$} &
   \colhead{DATE-OBS} &
   \colhead{$t_{\rm exp}$} &
   \colhead{Instrument} \\
    \colhead{} &
    \colhead{} &
   \colhead{(hh:mm:ss)} &
   \colhead{(dd:mm:ss)} &
   \colhead{(mag)  } &
    \colhead{YYYY-MM-DD} &
   \colhead{(s)} &
       \colhead{}  } 
\startdata
Collinder110  & 1103  & 06:38:39.61    & +02:05:49.51  &  15.0  & 2012-02-05   & 280  & MIKE \\
Collinder110   & 1120  & 06:38:45.90    & +02:05:59.83   & 14.0  & 2012-02-05   & 120  & MIKE \\
Collinder110   & 1122  & 06:38:49.18    & +02:06:31.15   & 13.7  & 2012-02-05    & 90  & MIKE \\
Collinder110   & 1128  & 06:38:50.12   &  +02:05:45.95   & 15.1  & 2012-02-05  &  500  & MIKE \\
Collinder110   & 1134  & 06:38:44.96   &  +02:04:23.66   & 13.7  & 2012-02-05    & 90  & MIKE \\
Collinder110   & 1135  & 06:38:46.22   &  +02:04:31.32   & 14.6  & 2012-02-05   & 200  & MIKE \\
Collinder110   & 1136  & 06:38:41.69   &  +02:03:51.70   & 14.9  & 2012-02-05   & 260  & MIKE
\enddata
\tablenotetext{}{Table \ref{tab:oc_starlist} is published in its entirety in the electronic edition of AJ, A portion is shown here for guidance regarding its form and content.}
\end{deluxetable*}

\subsection{Collinder 110}
Using the \cite{2003MNRAS.343..306B} photometric catalog, we
 selected RG and SG candidates with a magnitude cut of $V< 15.2$ (for both samples)  and color cuts of $B-V \geq 1.2$ and $0.98 \leq B-V < 1.2$, respectively. 
The list of RG candidates in WEBDA includes 71 stars, and this list contains all six of the photometric SGs and 41 of the 45 photometric RG candidates. 
All of the photometric candidates were observed with the exception of a single SG candidate. The observed stars are listed in Table \ref{tab:oc_starlist}, adopting the number system of \cite{Dawson:1998hy}.  Table \ref{tab:oc_starlist} lists the stars' coordinates, $V$ magnitudes, date of observation, exposure times, and instrument used.
 The adopted cluster age comes  from WEBDA, which is originally from \cite{Dawson:1998hy}.  The adopted reddening and distance modulus come from \cite{2003MNRAS.343..306B}, while the  [Fe/H]  comes from \cite{2010A&A...511A..56P}.

\subsection{Melotte 66}
Melotte 66 is a rich cluster that was observed over two different runs. 
We photometrically identified over 200 RG candidate stars using the \cite{zloczewski07} photometric catalog. The selection criteria were $V \leq 17.0$ and $0.9 \leq V-I_{\rm c} \leq 2.0$.
Because of the number of red giant stars and the distance to this cluster, we did not identify any subgiant candidates. WEBDA provides a list of 245 RG candidates. 
The only coordinates found in WEBDA are J1950.0 equinox, and we precessed these to J2000.0 coordinates using the IDL routine {\it jprecess} in ``The Astronomy User's Library''\footnote{http://idlastro.gsfc.nasa.gov/}. 
These coordinates were cross-referenced with the Fourth United States Naval Observatory CCD Astrograph Catalog (UCAC4, \citealt{Zacharias:2013cf}), and the J2000.0 UCAC4 coordinates are provided in Table \ref{tab:oc_starlist}.
All coordinate matches were within 4\arcsec\ of those calculated from the precessed coordinates.
In February 2012, we observed 76 stars: 66 WEBDA and CMD-identified candidates,  9 identified by CMD only, and 1 WEBDA-identified only.
An additional 17  stars (both WEBDA and CMD-identified) were observed in May 2012. 
The numbering system for Melotte 66 follows \cite{1997AJ....113.1723K}.

Table \ref{tab:oc_list2} lists two sources of cluster properties: WEBDA and \cite{Salaris:2004gh}. Similar metallicities are found for both, but the ages differ  by almost a factor of two. The WEBDA values are adopted because \cite{Salaris:2004gh} do not provide self-consistently derived reddening or distance modulus.  More recently, \cite{sesito08} derived [Fe/H]$\sim-0.33$ from high signal-to-noise (S/N) observations of 5 red giant members.

\subsection{M67}

 M67 is the most well-studied cluster in our sample with numerous photometric catalogs available. It is  also the only cluster for which spectroscopic \vsini\ measurements are available in WEBDA for any of the RGs.  An initial selection of RGs and subgiants was made with the following cuts on the $V$ versus $V-I_{\rm c}$ photometric data from  \cite{Montgomery:1993kn}:  RG candidates have $V < 12.5$ and $V-I_{\rm c} >0.95$, and 
SG candidates have  $V-I_{\rm c} > 0.75$ and $13.1 < V \leq 12.5$.
Excluded from this initial target list are stars with membership probabilities $<50\%$ \citep{sanders77} and known spectroscopic binaries with $P<20$~days. 
Finally, we also excluded stars that had previously determined \vsini\ measurements.  These \vsini\ measurements come from \cite{2001A&A...375..851M} and \cite{2011A&A...527A..94C}, and all of these have \vsini$<3.5$~\kms.
The list of the 30 observed targets are given in Table \ref{tab:oc_starlist}.  
  The adopted star numbering system comes from \cite{Fagerholm:1906tm}.
A few of the stars we observed are known spectroscopic binaries,
including  stars 137, 236, and 240.
An additional binary candidate is 136 \citep[][S1072 in that paper]{2012A&A...545A.139P}. 

Despite how well studied M67 is, its reported age still varies by a factor of two in the literature. WEBDA gives an age of 2.56~Gyr, while \cite{Salaris:2004gh} gives an age of 4.3 Gyr. Both of these values are listed in Table \ref{tab:oc_list2}. 
\cite{2011A&A...527A..94C} notes that age estimates often range from 3.5--4.5 Gyr.  A more recent study by \cite{2012A&A...545A.139P} plotted older isochrones to a binary-cleaned sample of members and  noted that while an age of  $\sim4$~Gyr is  a reasonably good match,  many stars still sit above the main sequence. Following the more recent studies, the older age from \cite{Salaris:2004gh} is adopted for M67 in this work.

\subsection{NGC 2477}
WEBDA provides a list of 86 red giant candidates for this cluster.  Using the photometry of \cite{1997AJ....113.1723K},  83 RG candidates were identified by  selecting $V < 13.0$ and $V-I_{\rm c}\geq 1.0$,  and 19 subgiant
star candidates were identified with $V < 13.0 $ and $0.75 \leq V-I_{\rm c} < 1.0$. There are  71 RG candidates in common between the WEBDA list and the photometric selection.  The 77 stars successfully observed in this cluster have the following break down in selection:  64 are both WEBDA and photometric RG candidates, 9 are only photometric RG candidates, 3 are WEBDA-only identified candidates,  and 1 is a subgiant. These stars are listed in  Table \ref{tab:oc_starlist} using the \cite{Hartwick:1974gg} numbering system. There are 12 known spectroscopic binaries  in our observed sample, most of which have periods longer than a few hundred days \citep{Eigenbrod:2004dx}. 

WEBDA lists an age of  0.7~Gyr, but more recent studies tend to use 1~Gyr (e.g., \citealt{Eigenbrod:2004dx}, \citealt{Bragaglia:2008fz}, and \citealt{Salaris:2004gh}).  Table \ref{tab:oc_list2} lists properties from both WEBDA and  \cite{Salaris:2004gh}, the latter of which is adopted in this work. 

\subsection{NGC 2506}
There are 46 RGs listed in WEBDA.  We removed six stars that had no  J2000.0 coordinates available, eleven stars with unknown or low proper motion membership probabilities \cite[$\leq 30\%$,][]{Chiu:1981ck}, and one star (3271) that that had a mean \vhelio\ inconsistent with the cluster velocity \citep{Minniti:1995wj}. 
The observed stars listed in Table \ref{tab:oc_starlist} have photometry from \cite{McClure:1981ch}, which also defines the numbering system.  We applied no photometric cuts. In Table \ref{tab:oc_list2}, three sources of cluster properties are listed: WEBDA, \cite{Salaris:2004gh}, and \cite{Mermilliod:2007kv}. All three give rather consistent ages and [Fe/H].  The \cite{Mermilliod:2007kv} values are adopted because the distance and reddening are slightly better matched to the \cite{McClure:1981ch} photometry.

\subsection{NGC 2660}
For NGC 2660, we used the list of red giant candidates from WEBDA, which lists 39 stars. Thirty of these candidates  have photometry available in the \cite{Sandrelli:1999hw} sample, but two candidates were removed because their colors put them on the main sequence turn-off.  We observed the remaining 28 candidates that have \cite{Sandrelli:1999hw} photometry.   One of the stars in the red giant list, star 660, has $V=11.1$ in \citet[][which is 1248 in that study]{Hartwick:1973gs}, but the coordinates in \cite{Sandrelli:1999hw} point to a fainter $V=15.4$~mag star. 
A brighter star just south of 660, which turned out to be 661 at $V=10.6$, was observed instead.

The adopted cluster parameters are those given in WEBDA, but it is worth nothing that \cite{Salaris:2004gh} finds this cluster to be younger and more metal poor. Both sets of cluster properties  are given in Table \ref{tab:oc_list2}.

\subsection{NGC 6005}
There is relatively little data on this cluster, with most of the measurements in WEBDA coming from the study by \cite{Piatti:1998gs}, whose numbering system is adopted here. 
A list of 19 RGs is available on WEBDA, which appears to be consistent with photometric cuts on the \cite{Piatti:1998gs} data of $V \lesssim 15$ and $B-V \gtrsim 1$.
The cluster properties in Table \ref{tab:oc_list2} are all from WEBDA.  No metallicity  has been reported for this cluster in the literature, and solar metallicity is  adopted for convenience. All 19 candidates were observed and are given in Table \ref{tab:oc_starlist}.

\subsection{NGC 6134}
The list of 25 RGs on WEBDA appears to be the same as that of  Table 4 of \cite{Claria:1992vv}, who used  photometry cuts of $V < 12.6$ and $B-V > 1.00$ to identify RGs, plus one additional RG (27).   The numbering system is based on \cite{Lindoff:1972up} and extended by \cite{Claria:1992vv}. 
However, \cite{Claria:1992vv} describe star  28 and 136 as ``definite non-members''  and 205 as a probable non-member; these three stars are excluded from our sample.
Star 27 has $B-V=0.94$ in the  \cite{Lindoff:1972up} study, but \cite{Kjeldsen:1991vc} found a $B-V=0.985$, much closer to the \cite{Claria:1992vv} RG cut.   The latter describe  27 as a  possible composite binary. 

\cite{Ahumada:2013cp} has a relatively new photometric study of this cluster, and they re-derived cluster parameters.  They note that despite knowing reddening and [Fe/H], they still cannot obtain unique age and distance measurements.  Their best values match the values available in WEBDA, and both are listed in Table \ref{tab:oc_list2}. We adopt the WEBDA values simply because it provides a single age instead of a range. 
Unfortunately, \cite{Ahumada:2013cp} do not identify individual stars so we cannot update our photometric cuts. 

\subsection{NGC 6253}
Photometric RG candidates were selected using the \cite{montalto09} $B$ and $V$ band photometry and the criteria that $12 < V \leq 15$ and $1 \leq B-V < 1.35$. This color-magnitude cut yielded 31 candidates.
There are 44 red giant candidates listed on WEBDA, 23 of which overlapped our photometric candidates. Of the remaining 21 giants in the WEBDA  list, 5 had very low membership probabilities \citep[$\leq 5\%$,][]{montalto09} and were not observed. We reviewed the $B-V$ photometry from \cite{bragaglia97}  and found that an additional four candidate giants in the WEBDA list did not satisfy our $B-V$ color requirements. The remaining 12 candidates were added to the observing list.
Positions are provided in \cite{bragaglia97}.
We observed  all 43 top priority candidates. 
The numbering system adopted by WEBDA is that of \cite{bragaglia97}, which then increments sequentially for objects introduced in
later studies.

\subsection{Pismis 18}
This cluster was also studied by \cite{Piatti:1998gs}, and we again adopt their numbering system. No red giant list is provided in WEBDA, so we selected giants with color-magnitude cuts of $1.4 \leq B-V \leq 2.1$  and $V \leq 15$,  yielding the 16 giants listed in Table \ref{tab:oc_starlist}. 
 The cluster properties in Table \ref{tab:oc_list2} are all from WEBDA, which appear to have come from  \cite{Piatti:1998gs}.

\subsection{Ruprecht 147}
Ruprecht 147 is a cluster that was recently well studied by \cite{2013AJ....145..134C}.  They identify eleven stars as RG stars in the $g'$ versus $g'-i'$ CMD, which were targeted in this study. 
The RG stars  successfully observed  are  CWWID \#'s 1, 2, 6, 9, 11, 15 and 19,  \citep{2013AJ....145..134C} as well  as CWWID 7 (accidentally added to the list, but it had the right  2MASS colors for an RG star).  In this study, the numbering system comes from \cite{2005A&A...438.1163K}. In this system, CWWID \#1, 2,  6, 7, 9, 11, 15, and 19 are \#247, 448, 420, 545, 119, 511, 388, and 319, respectively. These targets are given in Table \ref{tab:oc_starlist}.
We adopted the age and reddening from WEBDA and the distance modulus and metallicity from  \cite{2013AJ....145..134C}.

\section{Observations and Data Reduction}
To be sensitive to low \vsini, this project requires the  high spectral resolution of an echelle spectrograph. Such instruments generally span the entire
optical spectrum, and the numerous lines available allow observations at very modest S/N per pixel. We  aimed for only S/N$\sim$10 per pixel.  
All of the stars listed in Table \ref{tab:oc_starlist} were observed over three runs with two telescope-instrument combinations at Las Campanas Observatory in 2012.  During three nights (beginning Jan.\ 28, 29, and 30)  on the  Ir\'{e}n\'{e}e  du~Pont telescope,  all targets in M~67 and NGC~2477 were observed with the echelle spectrograph.   During two nights (beginning Feb.\ 4 and 5) on the Clay telescope, all stars in Collinder~110 and most of the stars in Melotte~66 were observed with the Magellan Inamori Kyocera Echelle (MIKE) spectrograph. Over another three nights (beginning May 7, 8, 9),  the remaining stars in Mel~66 and all of the stars in the remaining clusters were observed with MIKE. 
In addition to the cluster targets, both radial velocity standard stars and stars with known \vsini\ were observed on every night. 

\label{sec:obs}

\subsection{\dupont\ Echelle data}
The detector for the du~Pont  echelle spectrograph is a SITe2K CCD with 24~$\mu$ pixels. 
The  instrument was set up with a 0.75\arcsec\ wide slit, which yields a spectral resolving power ($R_\lambda =\lambda/\Delta\lambda$) of 40,000 and spans a wavelength range of 3600--10100~\AA\ over 60 orders.
The data were reduced using standard packages and tasks in IRAF.  The raw image frames were bias and overscan corrected. The pixel-to-pixel sensitivity variations were removed using flat fields constructed from nightly quartz lamp observations. A two-dimensional scattered light model was subtracted from each image and cosmic rays were removed. Given the low S/N acquired in the cluster stars, we used a bright RV standard star spectrum (S/N$\sim$60--100 per pixel) to define the order trace. Thorium-Argon spectra taken every half hour provide the wavelength calibration. A fourth-order polynomial was used to fit and remove the blaze function in each order.  The resulting spectra were left in multi-dimensional format, i.e., they were  not stitched to make a one-dimensional spectrum.

\subsection{MIKE data}
The MIKE spectrograph has both red and blue channels that can be used independently.  Although data were collected in both channels, only the spectra from the  red channel  are used in this analysis. The detectors for each arm have a 2k$\times$4k format with 15~$\mu$ pixels, and the red data were binned $2\times1$ on chip (i.e., a factor of two in the spatial dimension).  
A slit width of 0.5\arcsec\ was chosen to attain  $R_{\lambda}\sim$~44,000.  The red data span a wavelength range of 4800--9400~\AA\  over 34 orders.

We used the Carnegie python pipeline\footnote{Available at \url{http://code.obs.carnegiescience.edu/mike}} developed for MIKE to reduce the data.   MIKE's echelle orders have significant curvature  compared to other echelle spectrographs, and the facility pipeline is better suited for this unique feature than standard IRAF tasks. The python pipeline completes standard CCD processing including overscan
subtraction, bad pixel masks, flat fielding, sky subtraction, extraction and wavelength calibration of the object spectra.  
Like the du~Pont data, the spectra were not collapsed to a one-dimensional format. 

\section{Measuring velocities}
\label{sec:measure}
Both rotational and radial velocities are measured by cross-correlating the stellar spectra of the cluster stars against the radial velocity standard stars observed on the same
night.  Generally, three individual RV standards are observed each night with 2--3 images taken of each star. 
We used our own IDL code for performing the cross-correlation, which has similar functionality to IRAF's \emph{fxcor} task.  The template and object spectra are both continuum normalized and sampled
on  a log-linear wavelength scale. 
Because the blaze function significantly reduces the S/N on the edges of each order,  we only used the middle 50\% of the pixels by ignoring 25\%  of both the leading  and trailing pixels.  
We excluded the orders that are significantly affected by strong telluric bands ($\lambda \approx$~6860--6890, 7170--7350, and 7600--7630~ \AA), and we cross-correlated the template spectra against each other
to test whether any other orders tended to cross-correlate poorly.
For the \dupont\ echelle  spectra, the orders used in the cross-correlation include  22--46, 51, 54, and 57--60.  The useable orders in the MIKE spectra are 39, 43--44, 46, 51--53, and 56--63. 

The peaks of the cross-correlation functions (CCFs) were fit with a 5-parameter Gaussian fit---3  parameters for the height, width, and center of the Gaussian, and 2 parameters to fit a linear background.
We used IRAF's \emph{rvcorrect} task to calculate the heliocentric radial velocity corrections needed to convert the relative RV offsets between the template and object spectra to 
heliocentric radial velocities (\vhelio) for the objects.   The full width at half maximum (FWHM) of the CCF peak was converted to \vsini\ using a calibration derived for each  template star on each instrument (see \ref{calib:vrot}).

Each order and template provides an independent measurement of \vsini\ and \vhelio\ for each star, so that over 100 independent measurements are made for each object.  However, given the modest S/N of our data, not all of these measurements are equally good. In some cases, the largest peak in the cross-correlation function may in fact be a noise peak. We first winnow the sample by removing cross-correlations for which the peak height is  less than 0.4. From the remaining measurements, we compute an average in the following manner.
First, we calculate a sigma-clipped mean of  \vsini\ and \vhelio\ independently using the ``resistant\_mean.pro'' function in the IDL Astronomy User's Library, using a sigma-cut of 2.5.  The function returns an array of the values that are {\it not} rejected.  By comparing the lists of kept values, we determine which CCFs were kept for both their \vsini\ and \vhelio\ measurements. The \vsini\ and \vhelio\ measurements from this subset of cross-correlations are averaged to get the final  \vsini\ and \vhelio\ for the object star.

The uncertainty in \vhelio\ for each star is the total quadrature sum of the individual errors in \vhelio\  from fitting  the cross-correlation peaks.  In almost all cases, simply computing the standard deviation in the \vhelio\ measurements gives nearly the same result.
The uncertainties in \vsini\ are more complicated to compute, especially at the lowest values of \vsini\ where errors in the FWHM measurement can lead to unphysical solutions, namely \vsini$<0$. 
This problem is especially prevalent when attempting to propagate FWHM errors that are calculated from the cross-correlation peak fitting. 
We decided that the best method  to determine the uncertainty was from the distribution of \vsini\ measurements for each star.
In addition to avoiding the problems of propagating FWHM errors, this method for calculating uncertainty should also account for any systematic differences in the \vsini\ measurements that arise from using different template stars. However, because  FWHM measurements that would result in \vsini$<0$ are treated as giving \vsini$=0$, the distribution of \vsini\ measurements for a given  slowly rotating star is distinctly non-Gaussian. 
Therefore, instead of calculating the standard deviation, we compute the cumulative distribution function to find the \vsini\ values that that are the equivalent of 1 $\sigma$ away from the mean.  For stars where the \vsini\ uncertainty exceeds the \vsini, we report an upper limit that is the sum of the two.

\subsection{Calibrating \vsini}
\label{calib:vrot}
The method for finding the translation between the measured FWHM of a cross-correlation peak and \vsini\ was generally the same for both instruments. 
We artificially broaden the spectra of the RV standard stars by known \vsini, cross-correlate the spectra with unbroadened templates, and measure the FWHM of the cross-correlation peaks. This allows us to identify the relationship between input \vsini\ (unknown in the cluster stars) and the FWHM that we measure. 

We pooled all of the RV standard stars for each instrument together, selecting one image of each star to serve as an ``object'' spectrum and the remaining images for each star to serve as the templates.
The object spectra were then rotationally broadened with a grid of \vsini\ ranging from 2--6~\kms\ in steps of 1~\kms\ and 6--26~\kms\ in steps of 2~\kms.  The broadening kernel was computed using the IDL routine ``lsf\_rotate.pro,''\footnote{Available in the Astronomy Users's Library at \url{http://idlastro.gsfc.nasa.gov/contents.html}}  which is based on Equation 17.12  in \cite{Gray:1992wb}, using a limb-darkening parameter of  $\epsilon=0.6$.
All of the object spectra (both the original and the set of rotationally broadened spectra) are cross-correlated with the unbroadened template spectra, using the same algorithm on the CCF centers and widths to identify only ``good''  correlations. 
We assumed an intrinsic rotational velocity of 1~\kms, which has been measured for HD~66141 and HD~107328 \citep{1999A&AS..139..433D}. 
A calibration was defined for each order of each template star  using the mean measured FWHM and the total \vsini\ (quadrature sum of the kernel and intrinsic rotation). 
A quadratic interpolation between the grid points of these relationships yields a \vsini\ measurement from the FWHM of the objects' cross correlation function peaks. 
The calibration was only measured using self-correlations, i.e, the artificially broadened spectra were correlated only with  other images of the same star. 

There are slight differences in the calibration process for each instrument.
Seven unique RV standard stars were observed with MIKE:
CD-43~2527, HD~107328, HD~171391, HD~196983, HD~146051,  HD~66141, and HD~176047.  One spectrum of each of these stars
was selected to be rotationally broadened. Another 33 spectra of these stars were used as the cross-correlation templates.
We computed FWHM to \vsini\ relationships for each echelle order of every template star and found that a simple 3-pt quadratic interpolation of these relationships was sufficient for calculating \vsini\ from the FWHM of the objects' cross correlation function peaks.

 With the \dupont\ echelle, we  observed only three unique radial velocity standards (CD-43~2527, HD~66141, HD~107328). One of each spectra was used as an object star, and the remaining 24 spectra were used as templates.
We again calibrate FWHM and \vsini\  both  order-by-order and template-by-template, but 
we found  that a 4-pt least-squared quadratic interpolation provided a better mapping from FWHM to \vsini\ for the \dupont\ data.

Finally, we cross-correlated the unbroadened ``object'' spectra with the template spectra  of all the other stars (i.e.,  self-correlations were excluded),  to measure the \vsini\ for each of the RV standard  stars. This analysis verifies that all the RV standards are slow rotators, generally with \vsini$<2$~\kms.  In Table \ref{tab:vsini}, the top seven rows list the RV standard stars, the \vsini\ measured in this work from spectra taken with the MIKE and \dupont\ echelle instruments, and  \vsini\ reported in the literature.  All of the  \vsini\ measured for the template stars in this work are consistent with the assumed intrinsic \vsini\ of 1~\kms\ within the uncertainties. 
\begin{deluxetable*}{lrrrr}
\tablecolumns{5}
\tablewidth{0pc}
\tabletypesize{\scriptsize}
\tablecaption{Comparison of Rotational Velocities \label{tab:vsini}}
\tablehead{
   \colhead{Star} &
   \colhead{\vsini\ (MIKE)} &
   \colhead{\vsini\ (duPont)} &
   \colhead{Lit.\ \vsini} &
   \colhead{Reference}  \\
    \colhead{} &
    \colhead{(\kms)} &
    \colhead{(\kms)} &
    \colhead{(\kms)} &
    \colhead{}
}
\startdata
CD-43~2527     &                 $<$ 2.2  & 1.7  $\pm$   1.5 & \nodata & \nodata \\%
HD107328        &                  $<$ 2.7 & 1.8    $\pm$ 1.1 & 4.0, 1.3 & 1, 2  \\%
HD146051        &                 2.1  $\pm$ 1.8 &\nodata  & \nodata & \nodata \\%
HD171391        &                 $<$ 3.1 &\nodata & 2.5 &  3 \\%
HD176047        &                 $<$   0.5 &\nodata & \nodata & \nodata\\%
HD196983        &                 $<$ 0.7  &\nodata & \nodata & \nodata\\%
HD66141          &                 $<$ 0.6  & 2.0    $\pm$ 1.2 & 2.5, 1.1  & 1, 2  \\%
\hline
G0827-16.3424      &   23.9 $\pm$ 1.8  &  \nodata &    23.9 & 4\\
G0946+00.48        &   11.9  $\pm$    0.9  &  \nodata&  12.3 & 4 \\
G1124-05.61        &    6.7 $\pm$  0.5  &    5.4  $\pm$   1.0 & 7.5 & 4\\
Tyc0319-00231-1    &     $<$ 1.5   &   \nodata &  0.7 & 4 \\
Tyc0347-00762-1    &   15.3 $\pm$  0.5  &    14.8 $\pm$  2.4  &  15.3 & 4 \\
Tyc0647-00254-1    & \nodata &  10.9 $\pm$  0.5  &    10.4 & 4 \\
Tyc5523-00830-1    &    $<$ 0.6  & \nodata   &     1.5 & 4 \\
Tyc5868-00337-1    &    \nodata  &   $<$1.8  &     2.2 & 4 \\
Tyc5981-00414-1    &    4.2  $\pm$   0.7  & \nodata &    5.6 & 4 \\
Tyc5904-00513-1    &   \nodata  & 13.9 $\pm$  0.9  &    14.0 & 4 \\
Tyc6094-01204-1    &   13.7  $\pm$   0.6  &  13.2  $\pm$   1.3 &  13.2 & 4
\enddata
\tablerefs{(1) \citealt{1997PASP..109..514F}, (2) \citealt{1999A&AS..139..433D}, (3) \citealt{2014A&A...561A.126D}, (4) \citealt{carlberg12}}
\end{deluxetable*}

\subsection{Verifying \vsini.}
This cross-correlation method implicitly assumes that the only broadening source that varies among the red giant stars is the rotation.  This assumption is valid for the instrumental broadening, which is identical for all stars observed on the same telescope with the same setup, and for the microtrubulence, which varies little across our sample. However, the macroturbulent velocity increases linearly with temperature and varies considerably across the RGB. From the base to the tip of the RGB, which spans a temperature range of $\sim$~3000~K, macroturbulence  should range from  $\sim 8.5$~\kms\ down to $\sim 3.5$~\kms. Most of the stars  in the sample ($\sim 87\%$) have temperatures in the range of 4500--5500~K, where macroturbulence only varies between 4.8--6.8~\kms. \citep[These macroturbulence values were calculated using the temperature relationships of][for luminosity class III giants.]{2007A&A...475.1003H}  However, the macroturbulent broadening most strongly affects the wings of the line profiles and should have less effect on the FWHM measurements than rotation. To test  whether the \vsini\ measurements are being affected by macroturbulence,  we computed the Spearman rank correlation coefficient ($\rho$) to test for a correlation between \vsini\ and stellar temperature.  The value of $\rho$ ranges from $-1$ to 1, with the extrema corresponding to perfect monotone anti-correlation or correlation and $\rho=0$ corresponding to no correlation.  We attained $\rho=0.09$, indicative of a weak positive correlation between  \vsini\ and temperature. 
This non-zero correlation is only significant at the 1.9$\sigma$ level. We conclude from the weakness of the correlation and its low significance that the differences in stellar broadening are most likely attributed to rotation and not macroturbulence.

\begin{figure}
\centering
\includegraphics[width=0.5\textwidth]{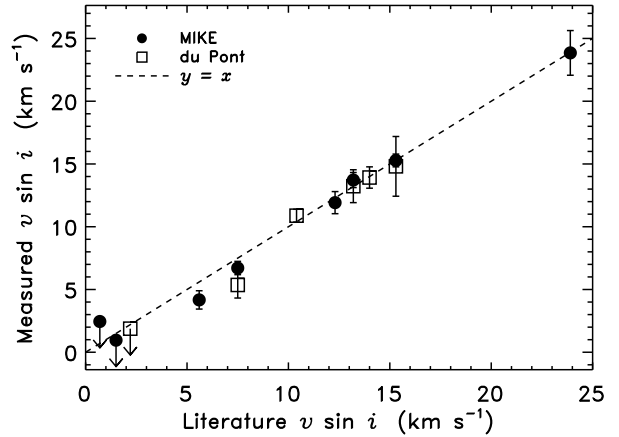}
\caption{\label{fig:vsini_check} Comparison between \vsini\ measured here and literature values for a sample of stars with known \vsini\ from \cite{carlberg12}. 
The different symbols refer to MIKE data (filled circles) and \dupont\ echelle data (open squares).}
\end{figure}

We can also test the validity of our technique by comparing \vsini\ for stars with previously measured rotation.
This comparison has already been made for the RV template stars used in the \vsini\ calibration, and we also observed a small number of stars  that span a wide range of \vsini\ specifically for this purpose.
These stars were selected from \cite{carlberg12} and were subject to the same reduction and analysis pipeline as the cluster stars. 
Eight of these stars were observed with MIKE, and they have  \vsini$\sim$0--24~\kms.  Only six of these stars were observed with the du~Pont echelle,  and they span a narrower ranger in \vsini, from $\sim$3--15~\kms. 
The stars are listed in Table \ref{tab:vsini}, and in 
Figure~\ref{fig:vsini_check}, we plot the  \vsini\ measured here against the literature values. A unity-slope line is drawn to guide the eye. 
Stars observed with MIKE and the \dupont\ echelle are shown separately since the \vsini\ calibrations were derived independently.  Both data sets reproduce the known \vsini\ values  very well, especially at higher \vsini, and there is excellent agreement for the three stars observed with both instruments.
However, it is worth noting that the three measurements at intermediate velocities, between 5 and 10~\kms, are systematically lower.
This rotation regime is where \vsini\ is low enough that macroturbulence becomes the dominant broadening, and suggests that we may be systematically underestimating \vsini\ by $\sim$~1.5~\kms\ at the lowest \vsini.

\section{Radial Velocity  Membership}
\label{sec:vhelio}
\begin{figure}
\centering
\includegraphics[width=0.5\textwidth]{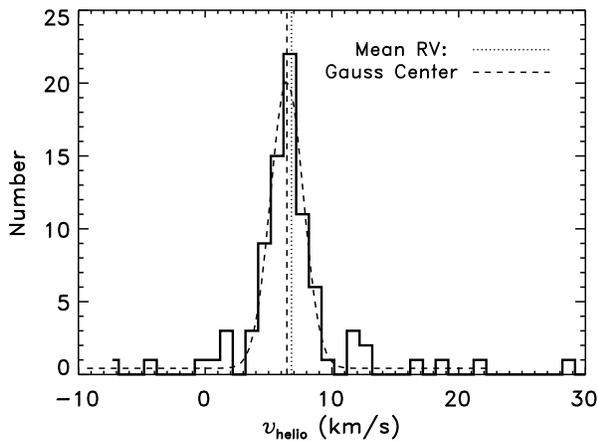}
\caption{\label{fig:some_cluster_RVs} Histogram of RVs  measured for RG candidates in NGC~2477 (solid). The dashed lines show  a Gaussian fit to the histogram and the location of the peak of fit. The  dotted line shows the mean RV of the likely cluster members. }
\end{figure}

We use the stars' \vhelio\ distribution both to measure the cluster's \vhelio\  and  to evaluate  cluster membership. First, we create RV histograms using a bin size of 1\,\kms, as 
illustrated in Figure~\ref{fig:some_cluster_RVs} for NGC~2477. This distribution is fit with a 4-term Gaussian function. The center of the Gaussian ($v_{\rm cen}$) gives the systemic velocity of the cluster, while the dispersion of  the Gaussian ($\sigma_{v}$)  is a measure of the intrinsic velocity dispersion within the cluster.  To be considered a cluster member, each individual star's \vhelio\ must be  within $v_{\rm cen} \pm 3\sigma_{v}$.
By defining $\Delta$ as the absolute difference between the stellar \vhelio\ and the mean cluster \vhelio,  this criterion can also be expressed as $\Delta/\sigma_{v} \leq 3$.
   Table  \ref{tab:oc_prop} lists each cluster along with the number of stars observed, the number of member stars, the clusters' \vhelio$\pm\sigma_{v}$  measured here,  and the literature \vhelio.  Because membership is here defined only  in terms of RV, the values for $\sigma_{v}$ should not be considered a robust measure of the clusters' velocity dispersions. 
Clusters that have been well-studied in the literature (e.g., M 67) have a high fraction of targeted stars being members. Less well-studied clusters (e.g., Melotte 66) have
a much larger fraction of field star contamination in the observed sample

\begin{deluxetable*}{lccccrrr}
\tablecolumns{6}
\tablewidth{0pc}
\tabletypesize{\scriptsize}
\tablecaption{Measured Properties of Open Clusters \label{tab:oc_prop}}
\tablehead{
   \colhead{Cluster Name} &
   \colhead{Observed} &
   \colhead{Members} &
   \colhead{$v_{\rm helio}$} &
   \colhead{Lit. $v_{\rm helio}$} &
   \colhead{Reference}\\
      \colhead{} &
   \colhead{} &
   \colhead{} &
   \colhead{(\kms)} &
   \colhead{(\kms)} &
   \colhead{}  }
\startdata
Collinder 110      &  51  &  29 &$+38.7\pm0.8$   &   $+41.0\pm$3.8& 1  \\
Melotte 66	      &  90  &  36  & $+22.1\pm0.9$   &   $+21.3\pm$0.4& 2  \\
M 67                  &  29 & 21    &  $+33.1\pm0.6$  &   $+32.0\pm$1.1&  3 \\ 
NGC 2477          & 81   &  65  &  $+6.9\pm1.2$   &   $+7.3\pm1.0$&  4  \\ 
NGC 2506	      &  28  &  24  & $+83.4\pm1.3$  & $+83.2\pm1.6$  &4 \\ 
NGC 2660          &  27  &  22  & $+21.8\pm0.8$  & $+21.3\pm1.0$  &  4  \\
NGC 6005	      &  18  &  11  &  $-25.2\pm0.8$ & \nodata & \nodata   \\
NGC 6134	      &  21  &  17  &  $-25.4\pm0.9$ &$-25.7 \pm0.7$   & 4    \\
NGC 6253	      &  40 &   26 &  $-28.2\pm1.2$& $-29.4\pm 1.3$  &   5 \\
Pismis 18	      &   16 &  12  &  $-27.9\pm0.8$ &  \nodata &  \nodata \\ 
Ruprecht 147     &  8  &   5   &   $+42.5\pm1.0$& $+41.1\pm0.5$  & 6 
\enddata
\tablerefs{(1)  \citealt{2010A&A...511A..56P}, (2) \citealt{sesito08}, (3) \citealt{2005A&A...438.1163K},  (4)  \citealt{Mermilliod:2008fw}, (5)    \citealt{2010AJ....139.2034A},
(6)     \citealt{2013AJ....145..134C}     }
\end{deluxetable*}

Table \ref{tab:oc_star_velocity} lists the stars' \vhelio, the uncertainty in \vhelio,  $\Delta/\sigma_v$ used to gauge cluster membership, \vsini, and the uncertainty in \vsini.
There is the possibility that legitimate cluster members will be classified as non-members with this strategy.  If the star has one or more stellar companions, the orbital component of the radial velocity can be large enough to make \vhelio\ inconsistent with the cluster velocity. Binary star systems
are discussed in more detail in \ref{sec:binaries}.

\begin{deluxetable*}{lrrrrrr}
\tablecolumns{7}
\tablewidth{0pc}
\tabletypesize{\scriptsize}
\tablecaption{Radial and Rotational Velocities \label{tab:oc_star_velocity}}

\tablehead{
   \colhead{Cluster } &
   \colhead{Star Number} &
   \colhead{$v_{\rm helio}$} &
   \colhead{err\_$v_{\rm helio}$ } &
   \colhead{$\Delta/\sigma_v$} &
   \colhead{\vsini} &
   \colhead{err\_\vsini} \\
    \colhead{} &
    \colhead{} &
    \colhead{(\kms)} &
    \colhead{(\kms)} &
    \colhead{} &
    \colhead{(\kms)} &
    \colhead{(\kms)} 
}
\startdata
Collinder110 & 1103 &  14.1 &  0.4 & 34.5 &  2.7 &  2.2 \\
Collinder110 & 1120 &  46.2 &  0.2 & 10.6 &  1.0 &    $<$ \\
Collinder110 & 1122 &  39.2 &  0.3 &  0.7 &  1.2 &    $<$ \\
Collinder110 & 1128 &  33.3 &  0.3 &  7.5 &  1.0 &    $<$ \\
Collinder110 & 1134 &  38.2 &  0.3 &  0.7 &  1.6 &  1.5 \\
Collinder110 & 1135 &  40.3 &  0.3 &  2.3 &  1.0 &    $<$ \\
Collinder110 & 1136 &  49.2 &  0.3 & 14.7 &  1.5 &  1.5 \\
Collinder110 & 1138 &  34.5 &  0.3 &  5.9 &  1.9 &  1.7 
\enddata
\tablenotetext{}{Table \ref{tab:oc_star_velocity} is published in its entirety in the electronic edition of \emph{AJ}, A portion is shown here for guidance regarding its form and content.}
\end{deluxetable*}

\section{Rotational velocities}
\label{sec:vrot_results}
\begin{figure*}
\centering
\includegraphics[width=\textwidth]{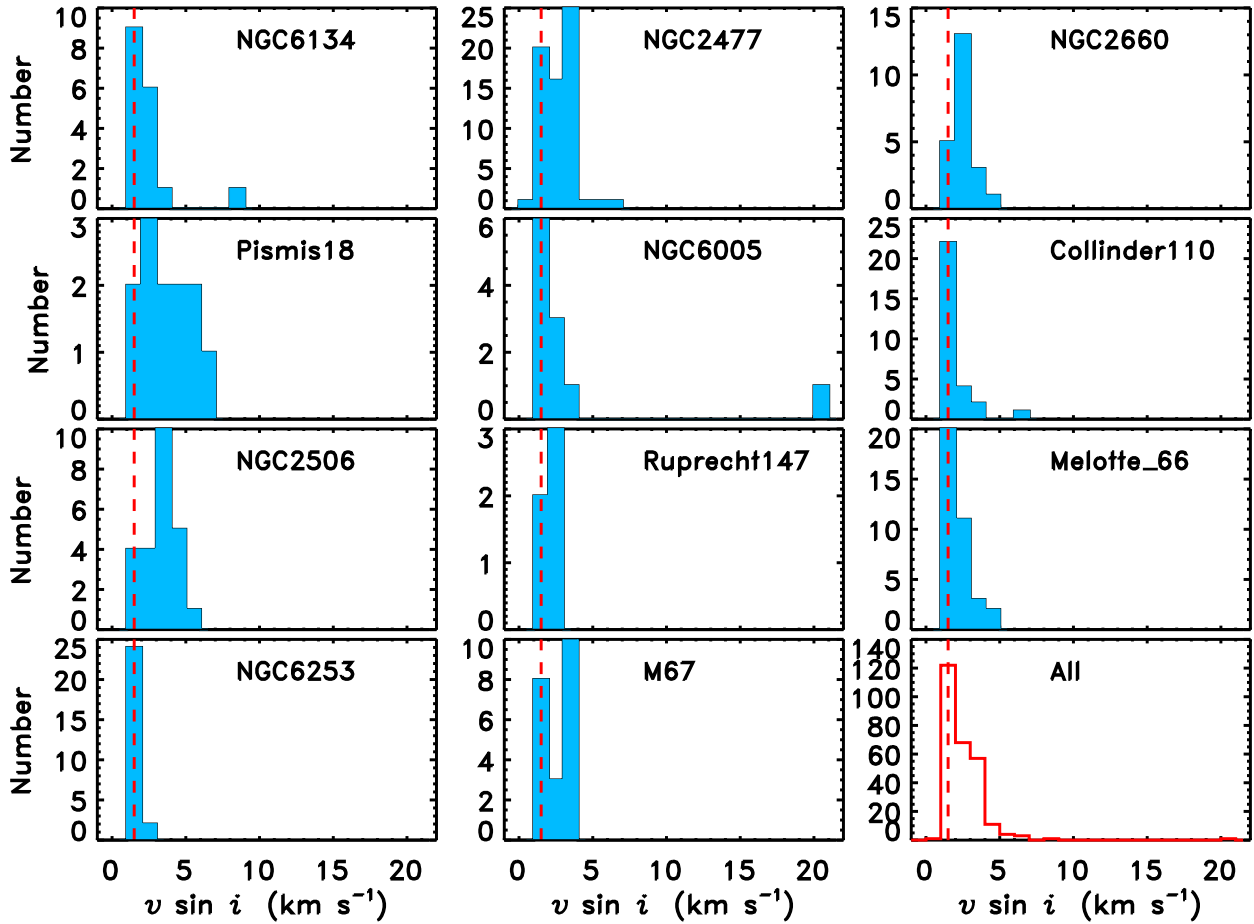}
\caption{\label{fig:vrot_dist} Distribution of \vsini\ for the individual clusters (filled) and for the total sample (open). For reference, the peak of the all-cluster distribution is show in each panel (dashed line). The panels are ordered from youngest to oldest (NGC 6134, NGC 2477 \dots M67). (A color version of this figure is available in the online journal.)}
\end{figure*}

In Figure~\ref{fig:vrot_dist}, we plot the distribution of \vsini\ among the member cluster stars, with the histogram for the entire cluster ensemble in the bottom right panel.  
The panels are ordered by the age of the cluster, from young to old. 
For ease of comparison, the location of the peak of the ensemble distribution (the bin centered at 1.5~\kms) is given in each panel.  The individual cluster distributions peak between 1.5~\kms\ and 3.5~\kms.
The distribution of the combined cluster sample shows that there is such a rapid decline in the number of stars with  increasing \vsini\ that stars with even moderate \vsini\ (4--5~\kms) are rare.   This result is consistent with previous studies of field RGs that generally find very slow rotation.
Only  two of the clusters have individual rapid rotators (defined here as \vsini$\geq 8$~\kms).  They are  in  NGC 6134 and NGC 6005.
Over half of the remaining cluster have at least one giant in  the still-rare range of \vsini\ that shows either moderate rotation (here defined as $6 \leq v \sin i < 8$~\kms) or modest
 rotation (here defined as $4 \leq v \sin i < 6$~\kms).  These clusters are NGC 2477,   NGC 2506, Pismis 18, Collinder 110, and Melotte 66. Pismis 18 in particular has a large fraction of such RGs.
All  cluster stars showing some level of enhanced rotation are listed in Table \ref{tab:oc_RRs}, and these are grouped by the three rotation bins described above (modest rotation, moderate rotation, and rapid rotation).  The notes in the last column are described in detail below.
\begin{deluxetable*}{lrrr}
\tablecolumns{4}
\tablewidth{0pc}
\tabletypesize{\scriptsize}
\tablecaption{Enhanced Rotators  \label{tab:oc_RRs}}
\tablehead{
   \colhead{Cluster Name} &
   \colhead{Star} &
   \colhead{\vsini} &
   \colhead{Notes\tablenotemark{a}} \\
   \colhead{} &
   \colhead{} &
   \colhead{(\kms)} &
   \colhead{}  }
\startdata
\multicolumn{4}{c}{Fast Rotation} \\
NGC 6005 & 4           & 20.0 & OL/CB?/RC/asymmetric CCF\\ 
NGC 6134 & 27         &   8.7 & SG, CB \\ 
\hline
\multicolumn{4}{c}{Moderate Rotation} \\
NGC 2477 & 6288      &  6.9 & OL\\[0.02in]
Pismis 18 & 41          &   6.5 & RC/AGB \\
Collinder 110 & 2119 & 6.3   &  RC \\ 
\hline
\multicolumn{4}{c}{Modest Rotation} \\
Pismis 18 & 22      	   &   5.7 &  RC/AGB\\
NGC 2477 & 4004      &  5.6 & OL \\
Pismis 18 & 2              & 5.1 & AGB \\
Melotte 66 & 2352      &  4.9 & RGB1  \\
Pismis 18 & 26          &   4.5 & AGB  \\
NGC 2506 & 2380      &  4.4 &RC, CCF is shallow, wingy. \\
NGC 2477  & 15496   &  4.2 & SG/RGBb/RGB1?\\ 
Melotte 66 & 745       & 4.1 & AGB/RC \\ 
Pismis 18 & 9             & 4.1 & AGB/RC 
\enddata
\tablenotetext{a}{CCFs are normal unless otherwise noted.  SG - suspected subgiant star, RGBt - tip of RGB, RGBb - base of RGB, RC - red clump, RGB1 - first ascent RGB, AGB - asymptotic giant branch (second ascent),  OL - over luminous star, CB - composite binary}
\end{deluxetable*}

\subsection{CMDs of Cluster Members}
\begin{figure*}
\centering
\includegraphics[width=0.33\textwidth]{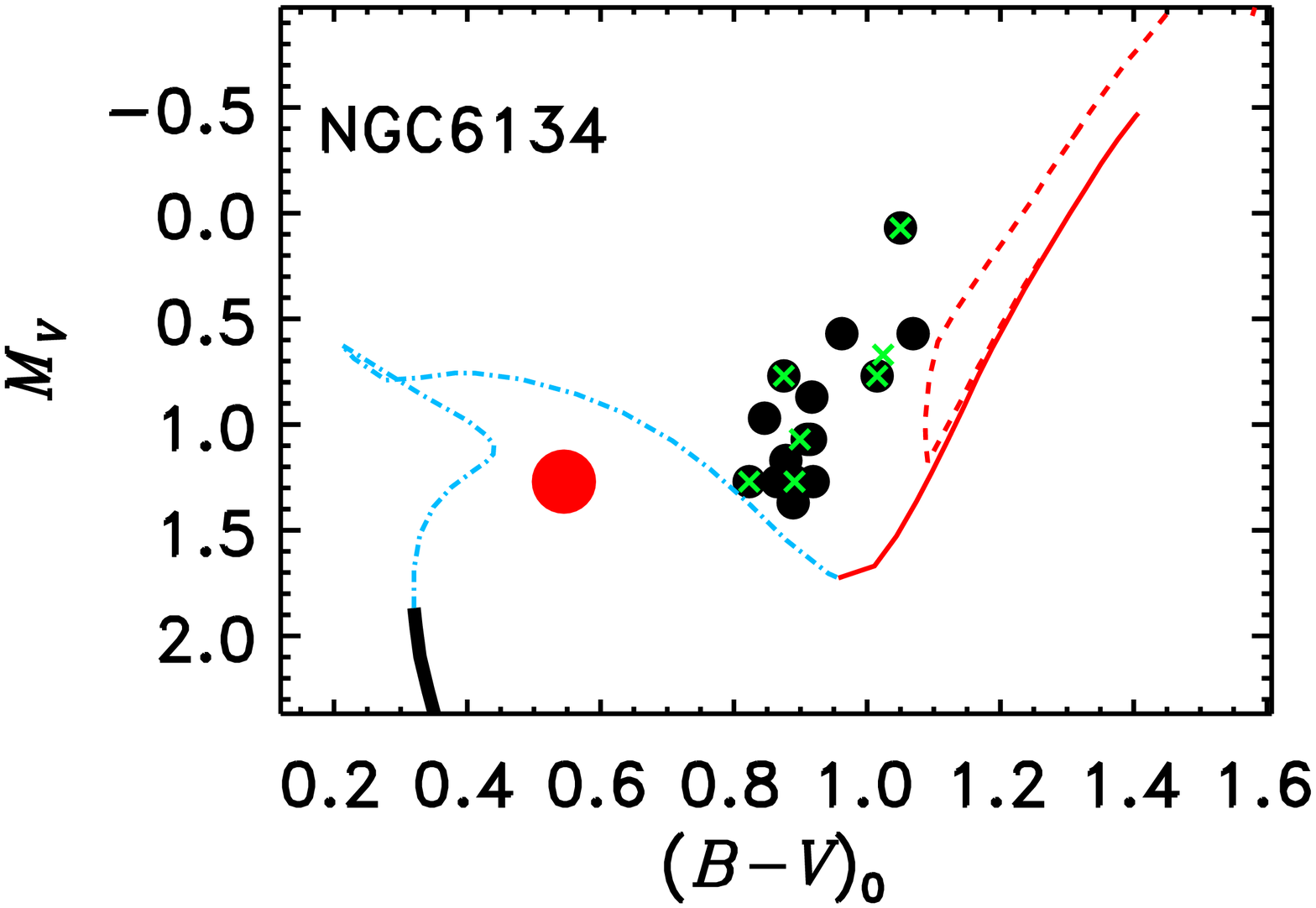}
\includegraphics[width=0.33\textwidth]{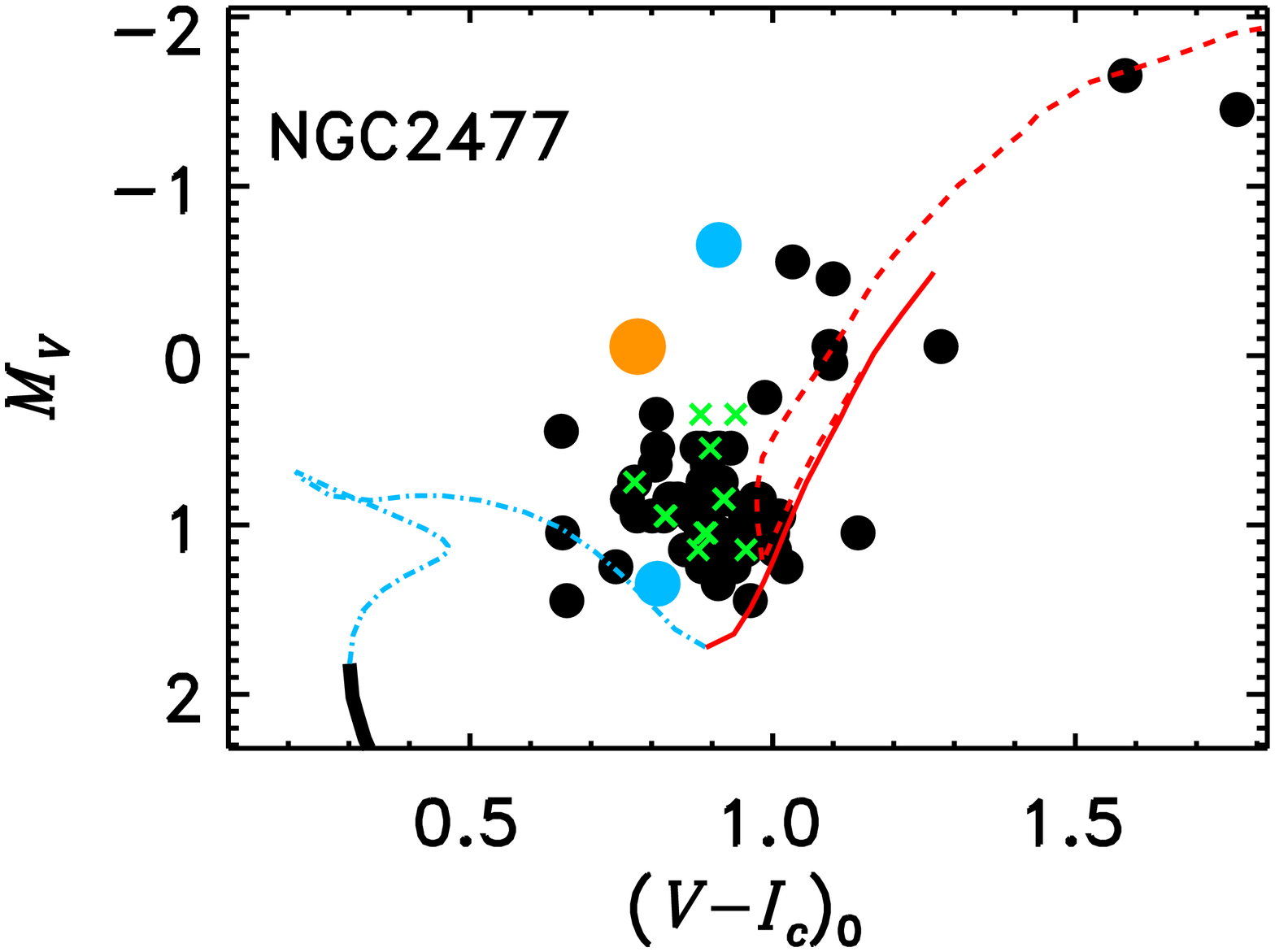}
\includegraphics[width=0.33\textwidth]{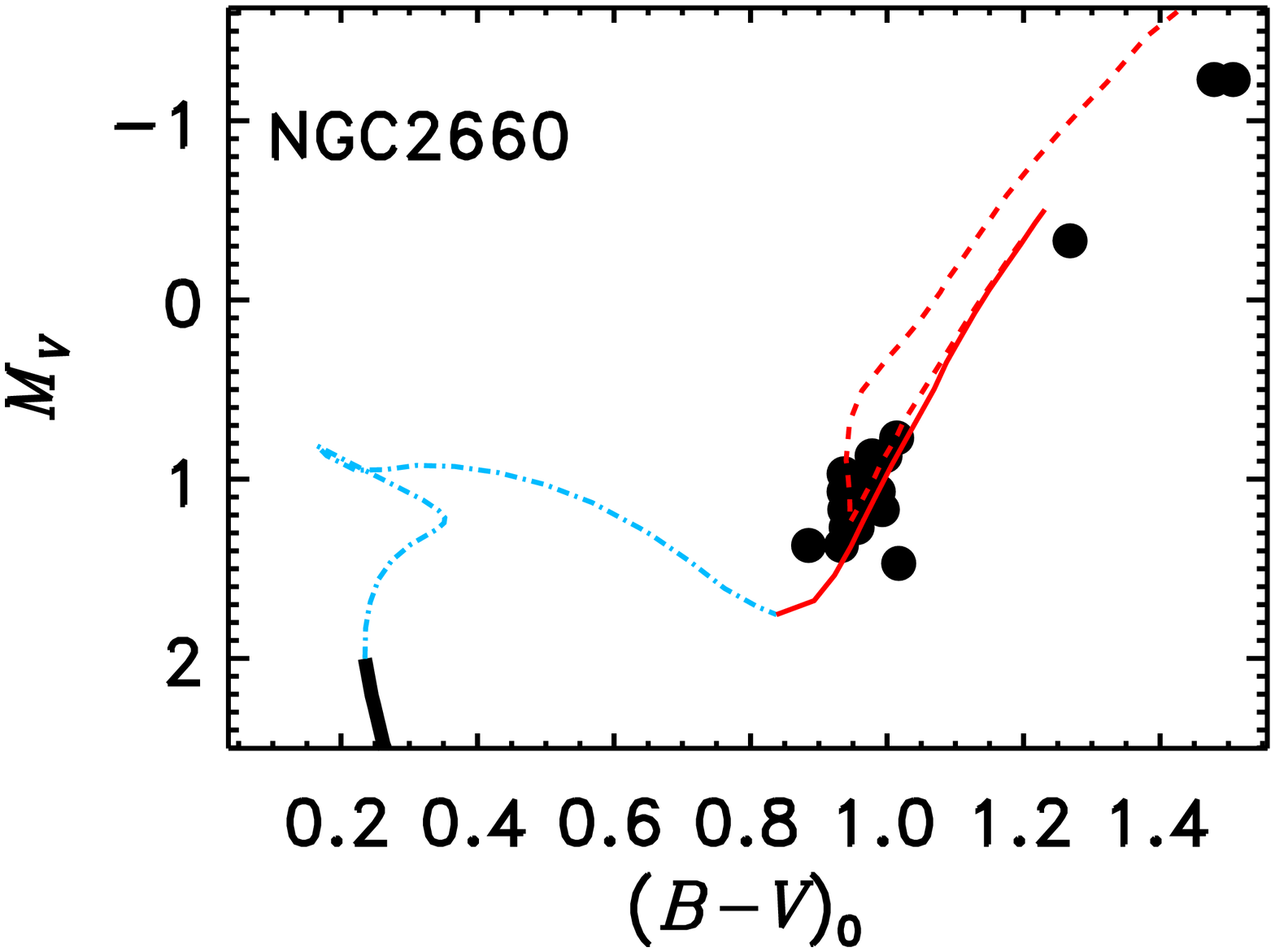}

\includegraphics[width=0.33\textwidth]{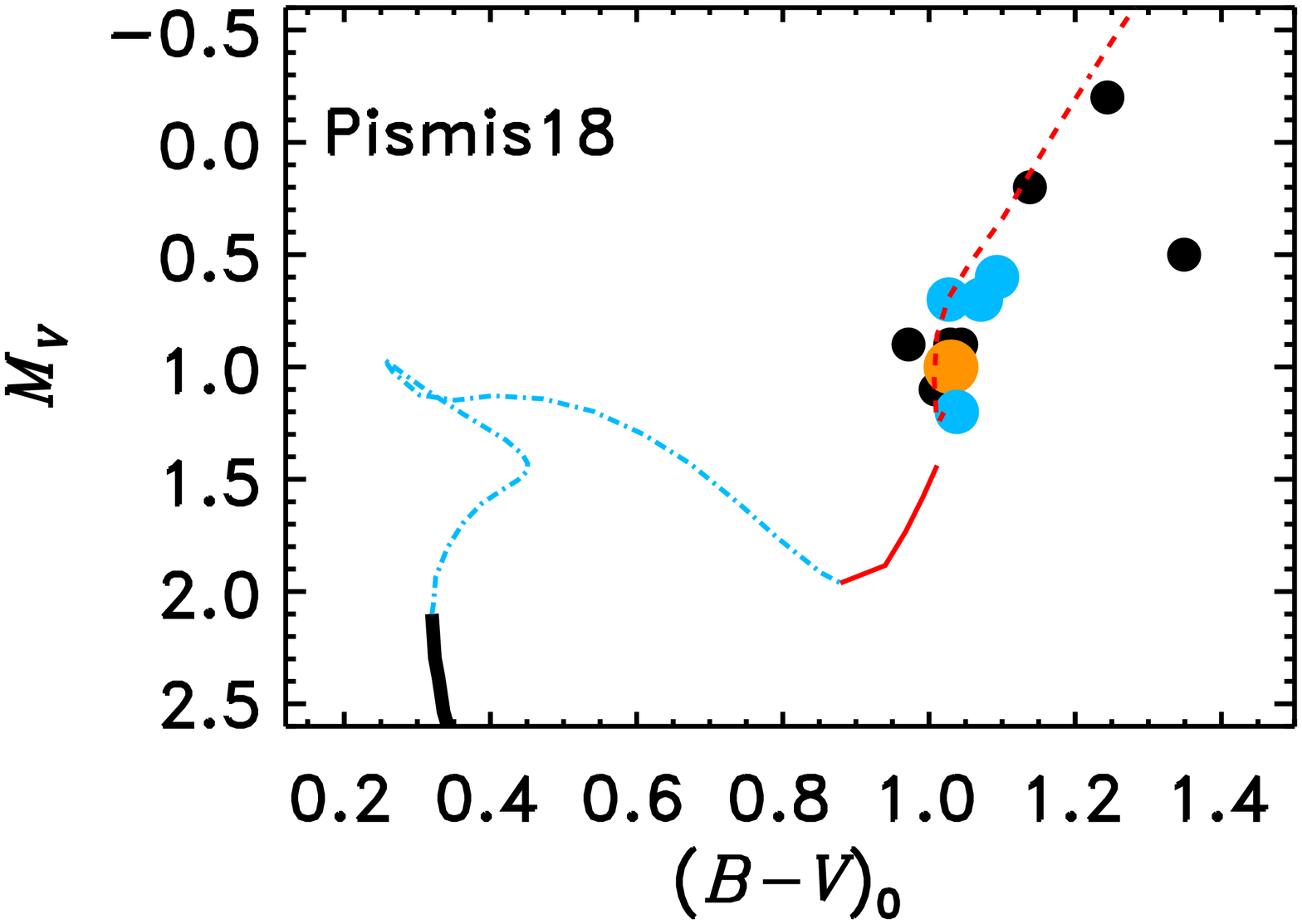}
\includegraphics[width=0.33\textwidth]{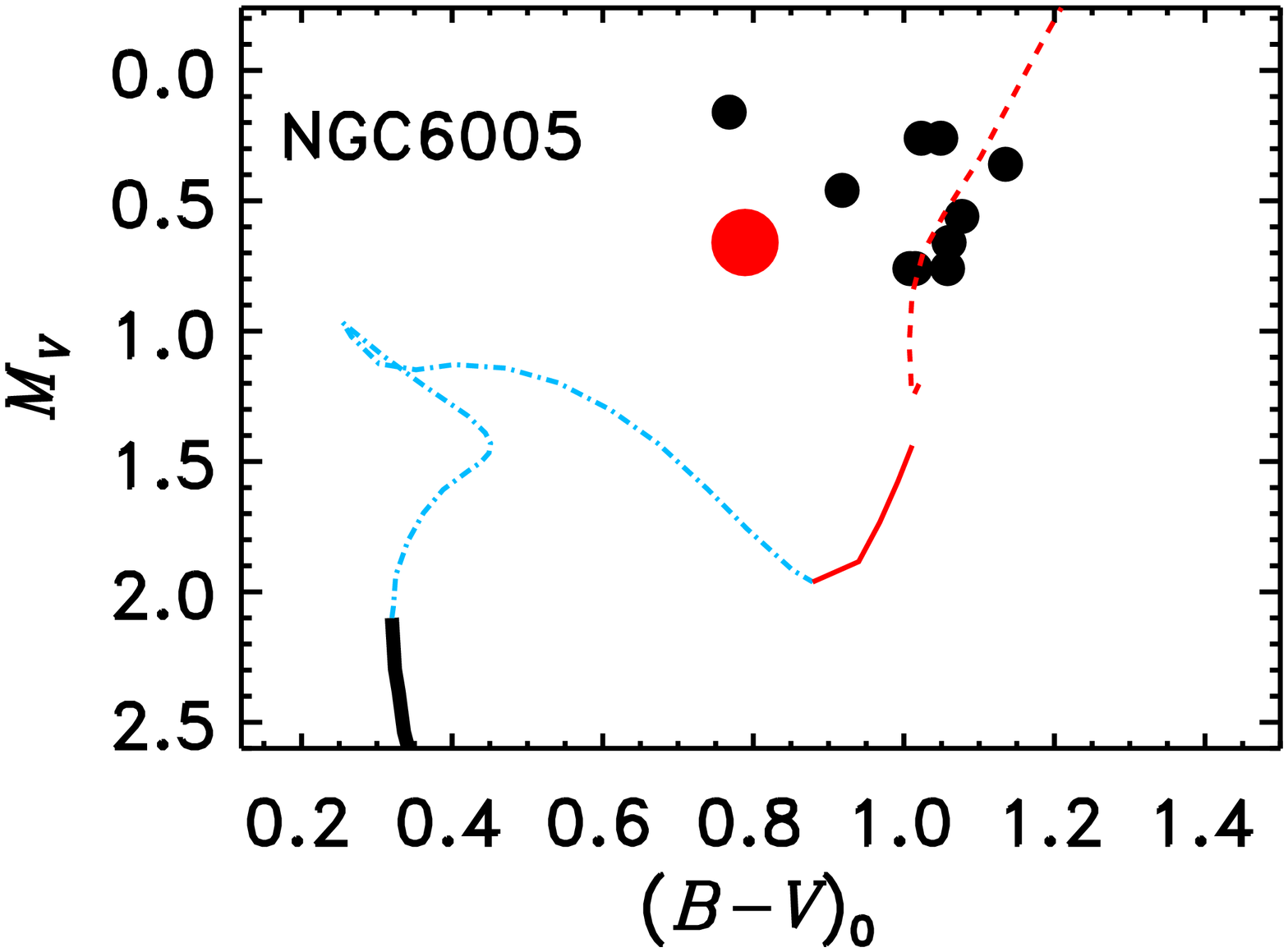}
\includegraphics[width=0.33\textwidth]{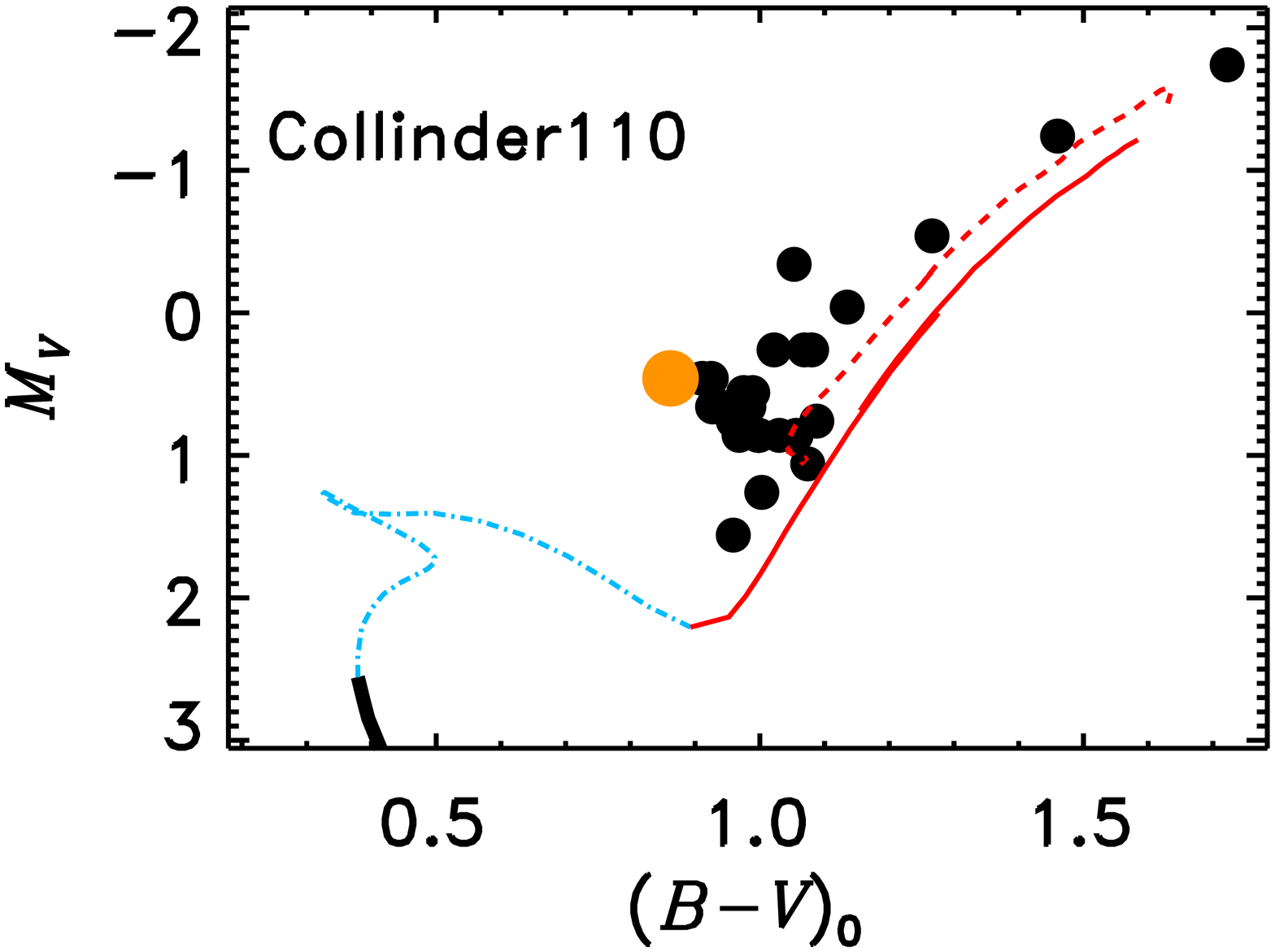}

\includegraphics[width=0.33\textwidth]{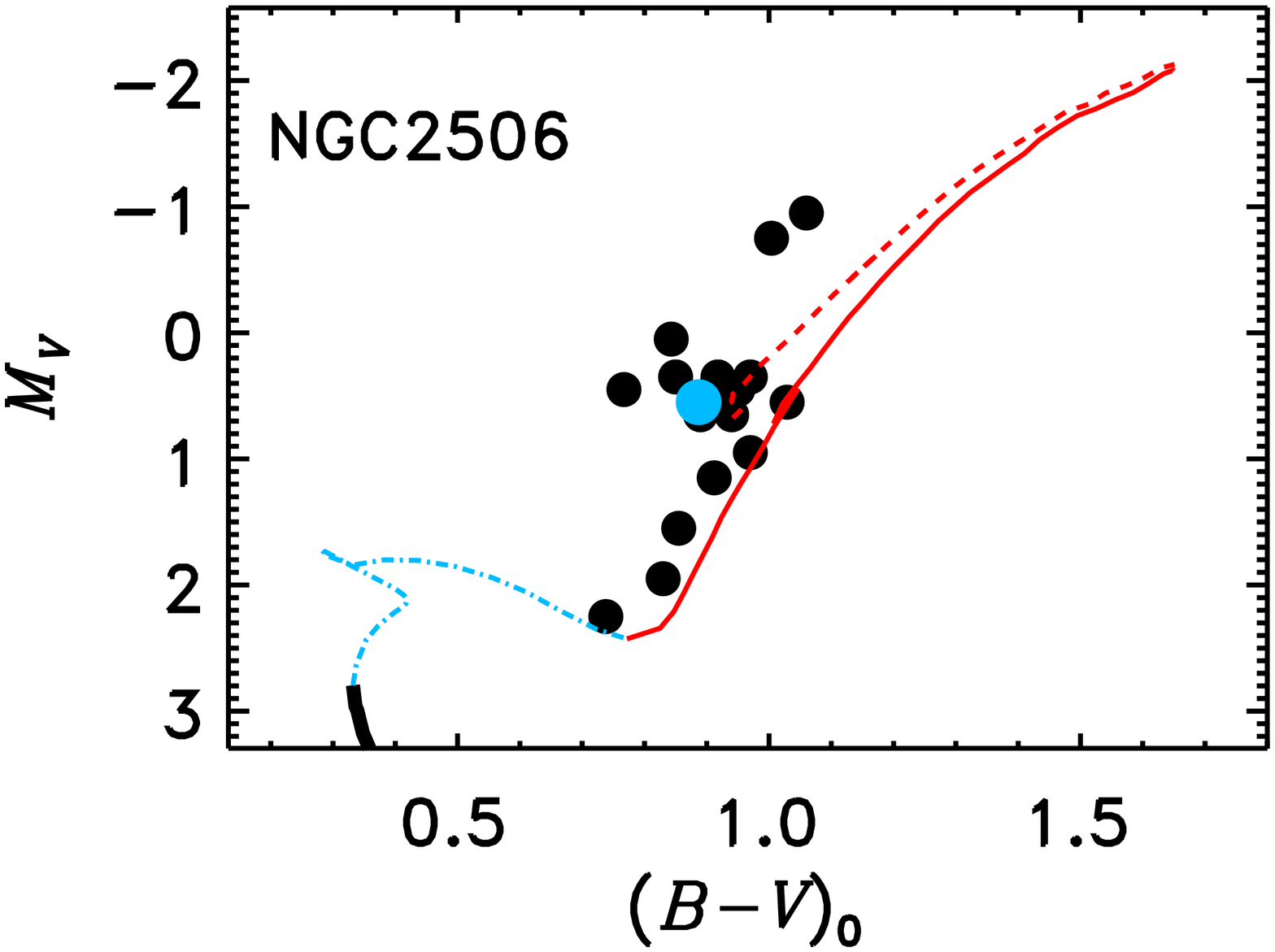}
\includegraphics[width=0.33\textwidth]{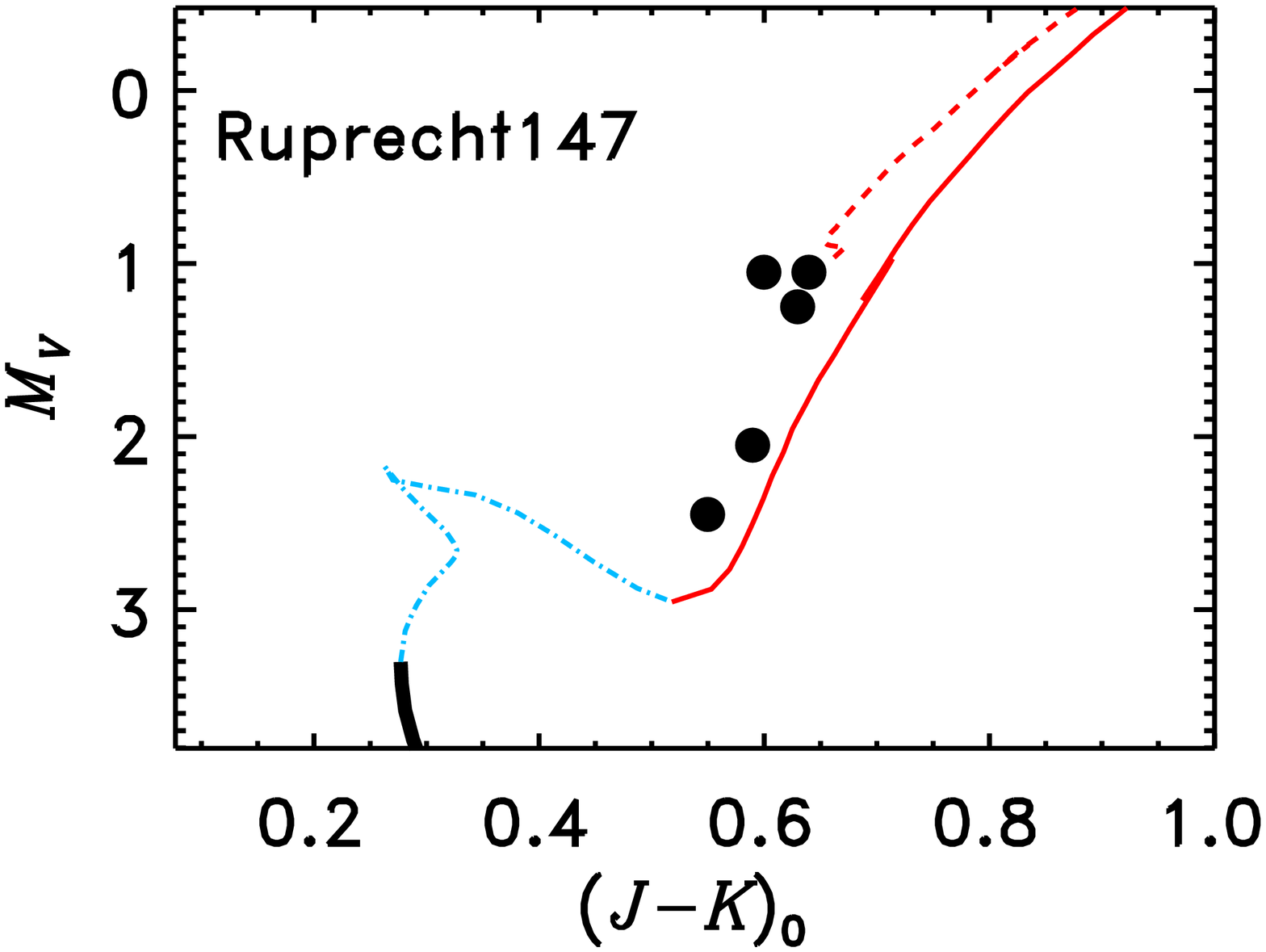}
\includegraphics[width=0.33\textwidth]{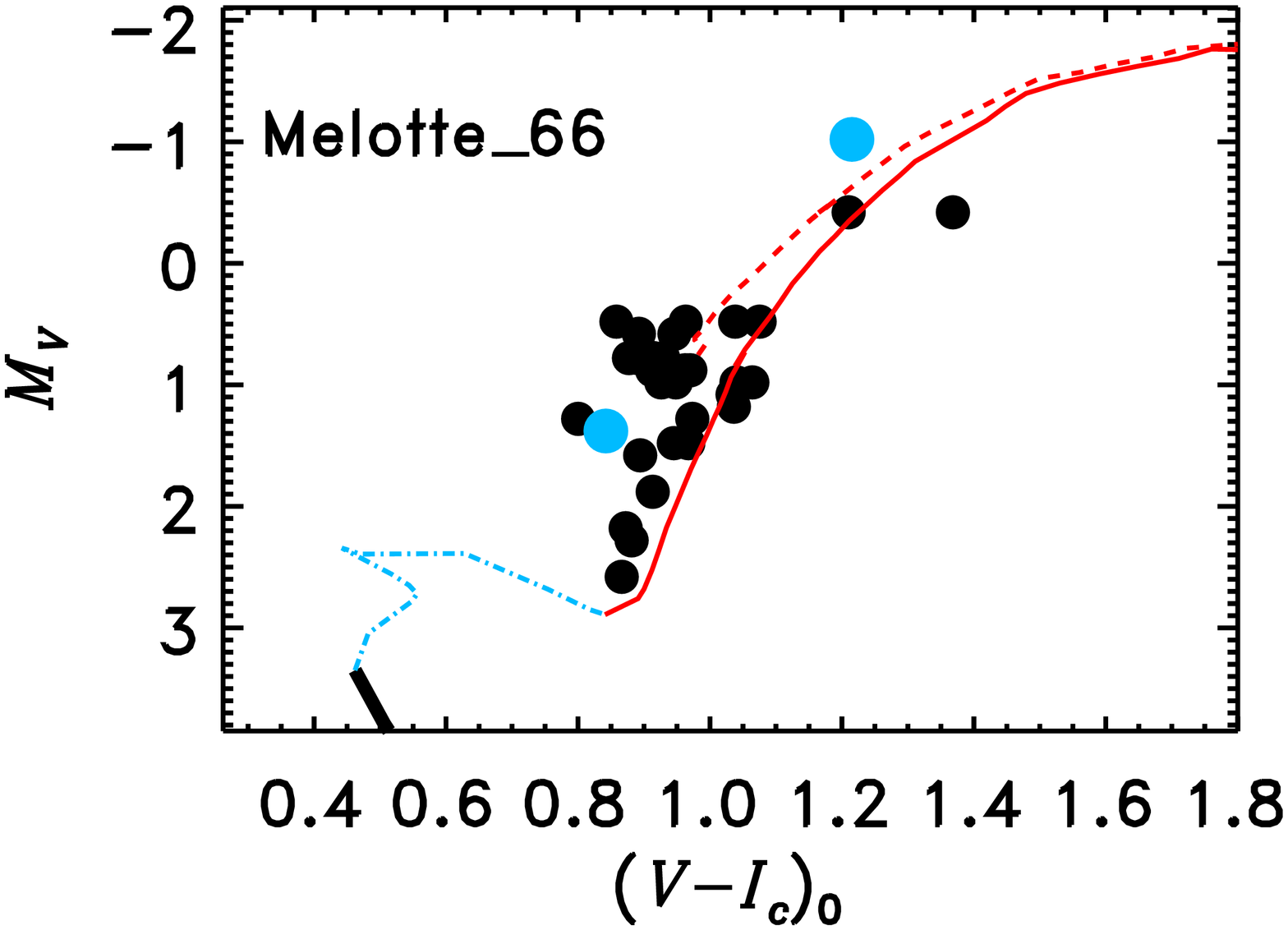}

\includegraphics[width=0.33\textwidth]{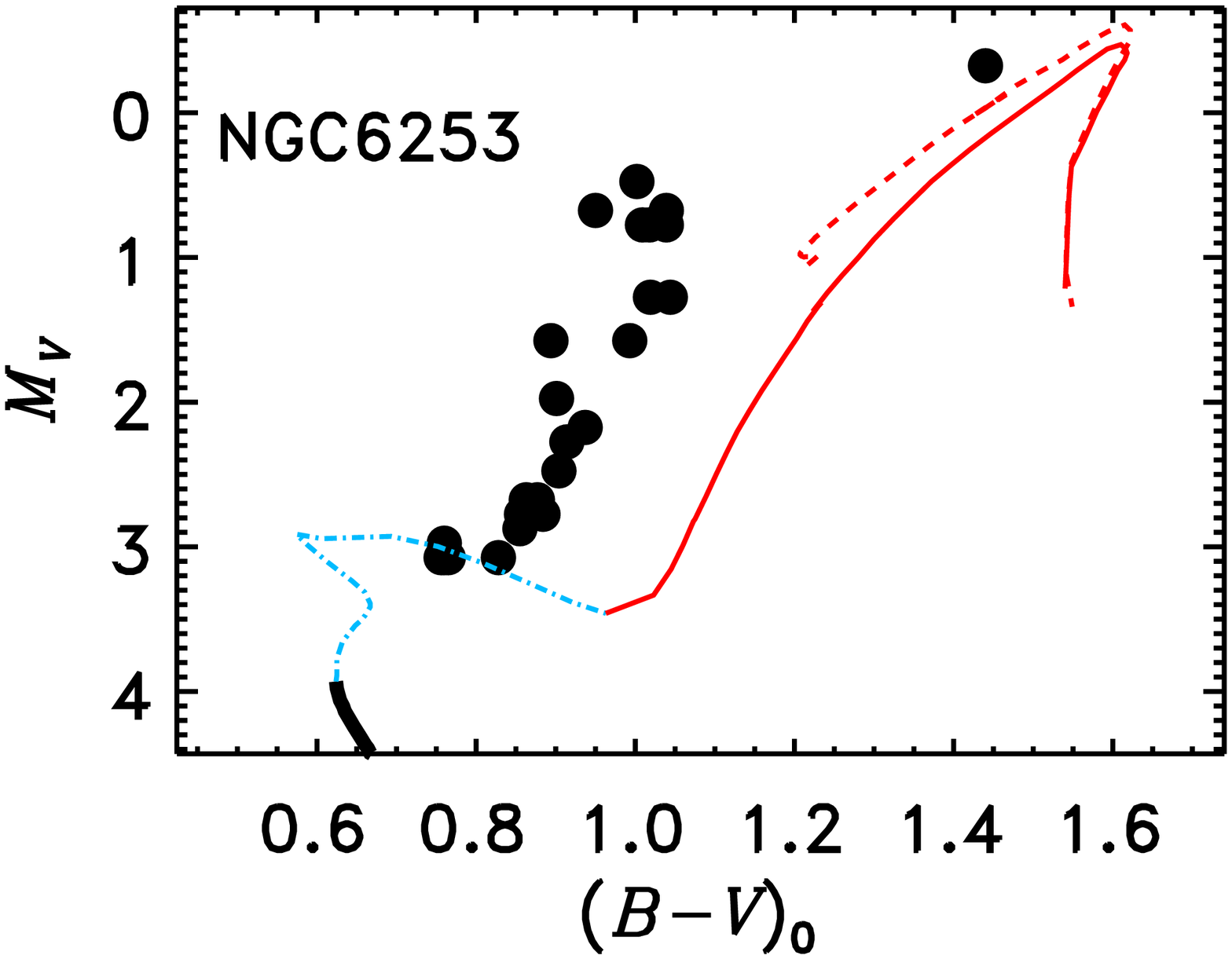}
\includegraphics[width=0.33\textwidth]{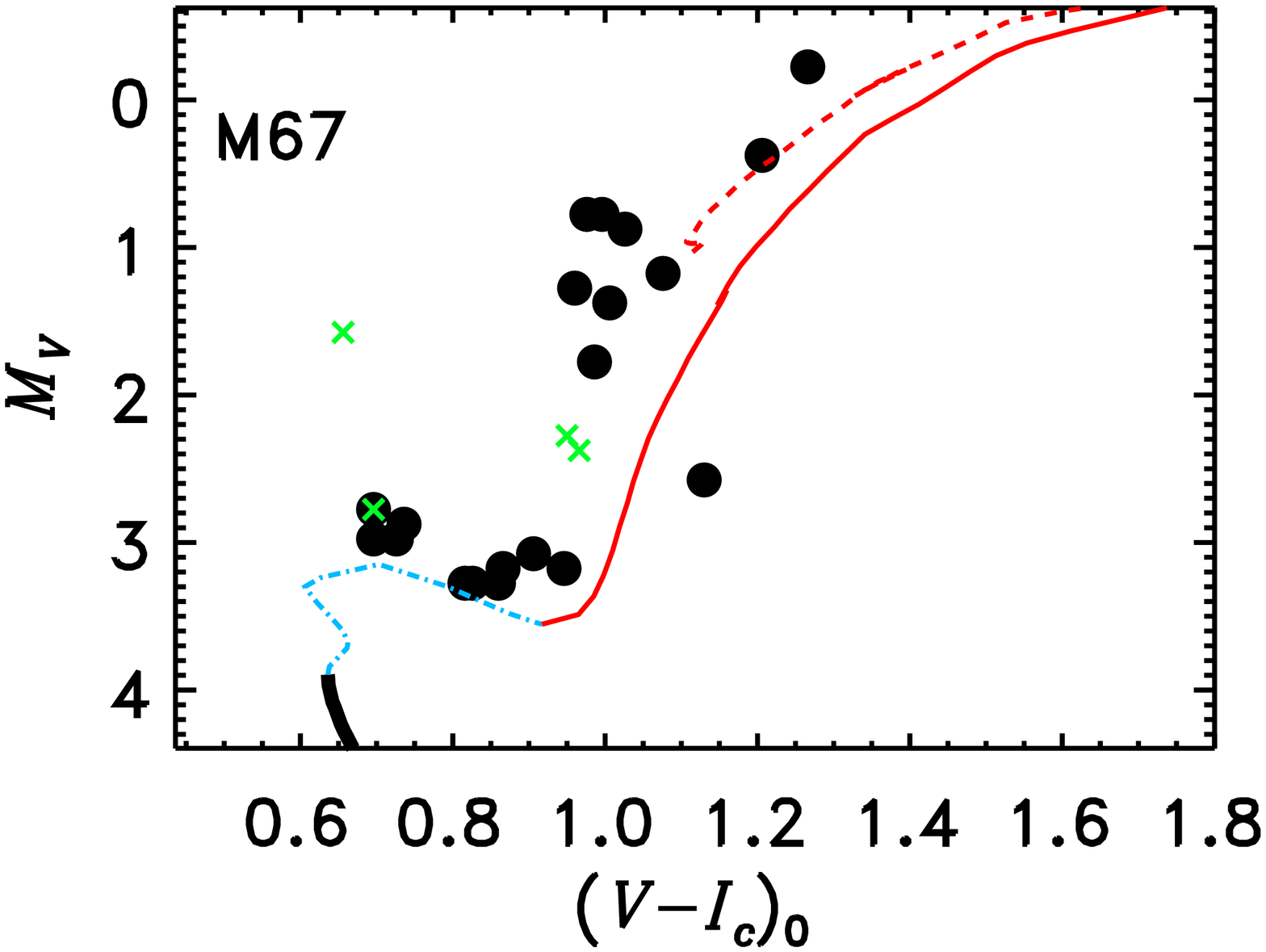}
\includegraphics[width=0.33\textwidth]{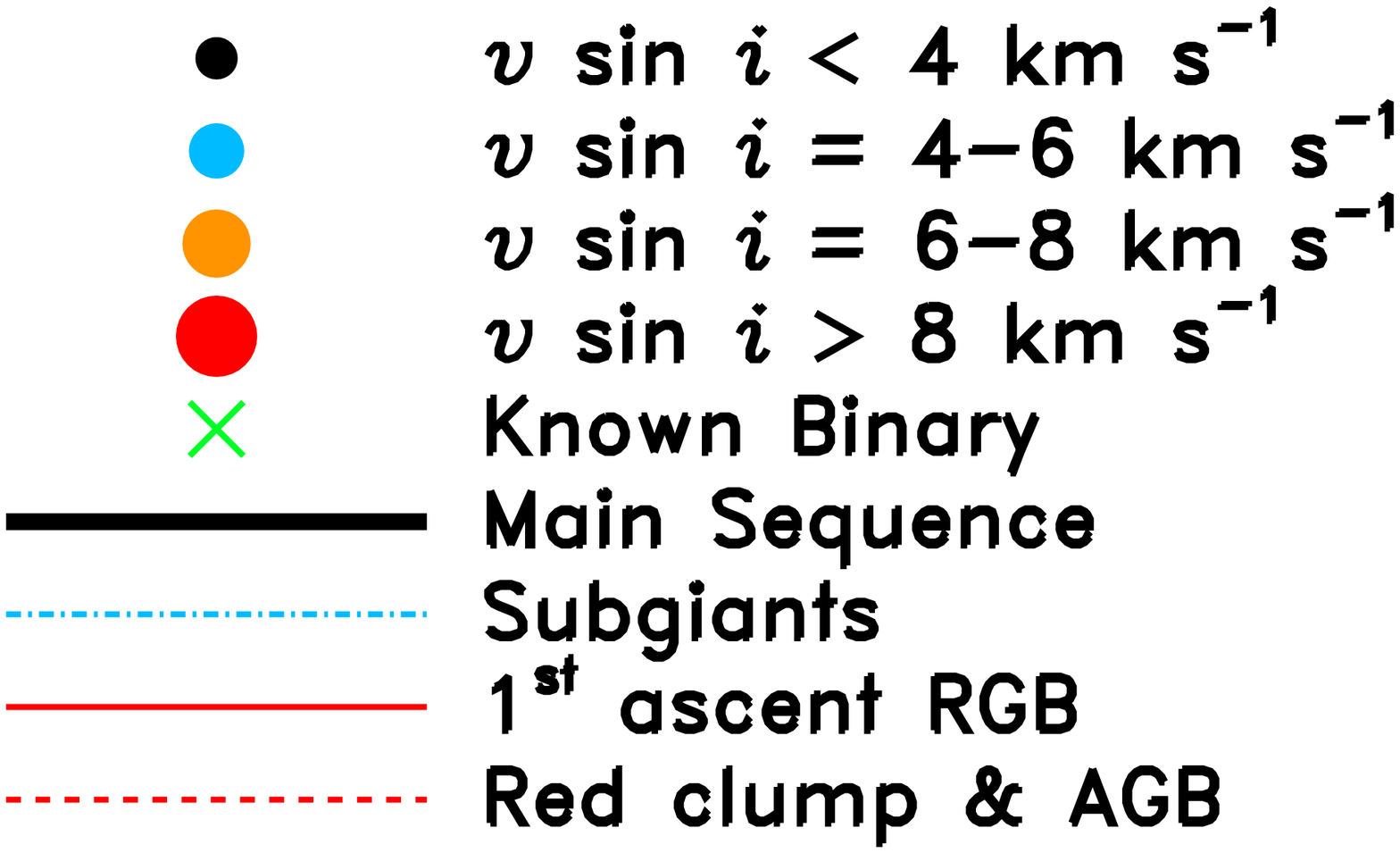}

\caption{Color magnitude diagrams of confirmed cluster members (circles) and model isochrones \citep{Marigo:2008fy} using the adopted cluster properties in Table \ref{tab:oc_list2}.  The  thick lines indicate the MS, and the  dashed-dot lines span the MS turn-off  and the base of the RGB.  The RGB evolution is shown with thin solid lines indicating the first ascent and dashed lines indicating the core He burning phase (red clump) and second ascent. The symbols are color coded and increase in size by velocity bins: slow ($v \sin i < 4$~\kms, black),  modest rotation ($4 \leq $\vsini~$ < 6$\,\kms, blue), moderate rotation ($6 \leq $\vsini~$ < 8$\,\kms, orange),   and rapid rotation (\vsini~$\ge 8$\,\kms, red).  The  $\times$'s in NGC~2477, M67, and NGC~6134 show stars that are known binaries.  (A color version of this figure is available in the online journal.) \label{fig:iso_rot}}
\end{figure*}

Before drawing any conclusions from these \vsini\ results, it is useful to see exactly where these stars fall on theoretical isochrones.  Figure~\ref{fig:iso_rot} shows  intrinsic color magnitude diagrams of all 11 clusters ordered again from youngest to oldest, together with \cite{Marigo:2008fy} isochrones of the adopted cluster ages and metallicities.  The $V$ magnitudes have been corrected for extinction using the standard extinction law, i.e., $A_V = 3.1 E(B-V)$. The $V-I$ colors for M67 and NGC 2477 were dereddened using $R_I=1.5$ \citep{Fitzpatrick:1999dx}\footnote{$R_\Lambda = A_\Lambda/E(B-V)$}, while the 2MASS $J-K_S$ colors for Ruprecht~147 stars were deredened using $R_J=0.819$  and  $R_{K_S}=0.350$  \citep{McCall:2004jh}.
No attempt has been made to fit isochrones to the data. The isochronal ages and metallicities and the clusters' distance moduli and reddenings  were chosen from the literature values in Table \ref{tab:oc_list2} with preference to sets that better matched the data. 

Most isochrones match the photometry quite well  considering they are not necessarily shown with the photometry from which the ages and metallicities were derived. There are two exceptions to this general statement.  
The especially poor fit to NGC~6253 is likely due to the fact that the most metal-rich isochrone available is at least 0.2~dex lower than the cluster metallicity.   Additionally, \cite{montalto09} shows that in $B-V$ photometry, their best-fit isochrone for this cluster fits the main sequence better than it fits the RGB.
The other cluster with an especially poor fit (NGC 6134) is the second most metal-rich cluster, and its metallicity is right at the metal-rich edge of the isochrone grid adopted in this work. \cite{Ahumada:2013cp} found it impossible to simultaneously fit the \vmic\  and $B-V$ isochrones to all the major evolution stages. 

The symbols in Figure~\ref{fig:iso_rot} are coded in size and color to represent four bins in rotation.  The slowest rotation bin (\vsini$<4$\kms) contains more than 90\% of the stars.  The remaining three bins in rotation  are those listed in Table \ref{tab:oc_RRs} and have  2--6 stars per bin.
The isochrone linestyles are coded to the following stages of evolution: the MS (heavy line), between the MS turn-off and base of the RGB (dot-dashed line), the first ascent of the RGB (thin solid), and the red clump/second ascent (dashed line).  From these CMDs, it is clear that the fast rotation is not always unusual. For example, in NGC~6134, the fastest rotator is clearly the least evolved star and is  still on the subgiant branch. The subgiants  are labeled `SG' in Table \ref{tab:oc_RRs}.  
This star  is only half way across the Hertzsprung gap (in temperature) and is still likely in the process of spinning down.   Contrast this star with the subgiants in M67, which  are all slow rotators already.

In two clusters, the scatter in the photometry makes it difficult to discern  the evolutionary stage of the stars with  enhanced rotation.  In NGC~2477, the scatter is likely due to differential reddening. \cite{Eigenbrod:2004dx} found that a single isochrone with reddening between 0.22 and 0.3~dex reproduced the photometry quite well.  Depending on the appropriate reddening value for the faintest moderate rotator, it may be  associated  with the subgiant branch, the base of the RGB or the red clump.
The other two rotators  are the brightest stars at their approximate color and are marked as over-luminous (OL) in Table \ref{fig:iso_rot}. However, the may simply be among the least-reddened stars in the cluster. 
 In NGC~6005,  the scatter around the isochrone is so large and the number of stars so few,  it is difficult to draw any conclusions from the rapid rotator's position on the CMD alone. It could either be associated with the red clump or be over luminous compared to the base of the RGB.

All of the remaining stars with enhanced rotation are in good agreement with some  red giant phase of evolution that is appropriate to the cluster, most commonly the red clump phase. 
 Curiously, Pismis 18 shows a strikingly high number of enhanced rotators: 5 out of only 12 members (over 40\%).  The isochrone is also the only one that does not have an ``RGB tip'' stage  (NGC~6005 uses the same isochrone, but its metallicity is unknown), leading to the apparent gap between the last stage  on the first ascent and the beginning of the core He burning stage. This is because the age of the cluster interpolates between isochrones with RGB stars that both do and do not have degenerate He cores.  
 All of the enhanced rotators are situated near the red clump.

\subsection{Binaries}
\label{sec:binaries}

\begin{deluxetable*}{lcrrrrrrr}
\tablecolumns{9}
\tablewidth{0pc}
\tabletypesize{\scriptsize}
\tablecaption{Known Binaries  \label{tab:oc_binary}}
\tablehead{
   \colhead{Cluster} &
   \colhead{Cluster \vhelio} &
   \colhead{Star} &
   \colhead{Period} &
   \colhead{$\gamma$} &
   \colhead{K} &
   \colhead{\vsini} &
   \colhead{Member?\tablenotemark{a}} &
   \colhead{Ref.} \\
   \colhead{} &
   \colhead{(\kms)} &
   \colhead{} &
   \colhead{(d)} &
   \colhead{\kms} &
   \colhead{(\kms)} &
   \colhead{(\kms)} &
   \colhead{} &
   \colhead{}  }
\startdata

 M 67      & 32.5  &  136 &   1495.0  &   32.87 &  2.60 & 13.2 &  M, N  & 1  \\   
              &           &  173  &   353.9   &  33.37 & 11.80 & 0.6 & M, N  & 1 \\   
              &           &  236  &   277.8   &  33.82 &  8.36 & 3.5 & M, Y & 1 \\    
              &           &  240  &  1233.0   &  43.30 &  4.40 & 0.4 & NM, N & 1 \\  
 NGC 2477 & 7.0 & 1025  &     41.554  &   6.72 & 11.75 & 3.7 & M, N  &   2 \\
                 &       & 1044 &    3108.0    &  7.46 &  3.22  & 0.8 & M, N & 2\\  
                 &       & 2064 &    4578.0    &  6.14 &  5.78  &  2.2 & M, Y& 2\\ 
                 &       & 2204 &    1318.9  &  7.62 & 19.47  & 1.6 & M, N & 2\\ 
                 &       & 3003 &    1782.4  &  7.49 & 11.13  &  1.7 & M, N & 2\\ 
                 &       & 3176 &     276.74  &  7.01 &  7.74  & 1.4  & M, Y& 2\\ 
                 &       & 4137 &     372.367 &  9.04 & 14.56  & 1.4 &  NM?, N & 2\\ 
                 &       & 5073 &     326.1    &  6.13 & 14.66  & 1.5 & M, Y & 2\\ 
                 &       & 6020 &     226.2    &  8.55 &  9.33  & 0.4 & NM?, N & 2\\ 
                 &       & 6062 &     482.3    & 10.16 &  6.21  & 0.2 & NM, Y & 2\\ 
                 &       & 6251 &     412.6    &  7.40 &  1.92  & 2.1 & M,Y & 2\\ 
                 &       & 8017 &      60.3169  &   8.33&  21.74 & 3.8 & NM?, N  &  2\\ 
                 &       & 8018 &     140.1490  &  6.28 &  6.77  & 2.2 & M, N & 2\\ 
NGC 6134 & $-24.9$ & 8  & 702.7 &  $-27.0$0 &  8.93  & 1.7 & M, Y    &3 \\
                 &          & 30  & \nodata   & \nodata  & \nodata  & 1.4 &   M, Y & 4 \\
                 &          &34  & 257.83 & $-26.12 $ &  8.87 & 1.4 & M, Y  & 3 \\
                 &          & 47  & \nodata   & \nodata  & \nodata    & 1.0 & M, Y  & 4 \\   
                 &          & 107 & \nodata   & \nodata  & \nodata    & 1.3 & M, Y  & 4 \\  
                 &          & 176 & \nodata   & \nodata  & \nodata    & 1.1 & M, N  & 4 \\  
                 &          & 204  & 59.674   & $-24.92$  & 9.50    & 6.7  & M, N & 3 
\enddata
\tablerefs{ (1) \citealt{1990AJ....100.1859M}, (2) \citealt{Eigenbrod:2004dx}, (3) \citealt{2007A&A...473..829M}, (4) \citealt{Claria:1992vv}}
\tablenotetext{a}{M/NM - membership from systemic velocity of system, Y/N - from \vhelio cut in this work. }
\end{deluxetable*}
Stars with close orbiting companions could have significant tidal interactions that affect the primary star's rotation.
Systems with the shortest period companions  are the most likely to have their rotations influenced by the companion. However,  these are also the systems where the primary has the largest orbital velocity variations, which  increases the likelihood of observing the system when the orbital RV component is large enough to give the star an observed RV that appears inconsistent with cluster membership.
We can use the clusters with previously identified binary star members to probe this likelihood.

In Table \ref{tab:oc_binary}, we list the binary systems in M67, NGC 2477, and NGC 6134, and these stars are plotted in Figure~\ref{fig:iso_rot} with $\times$'s.
The table gives the cluster name and star number used in this study; the \vhelio\ derived here; the binary systems' period, systemic velocity ($\gamma$), and velocity semi-amplitude ($K$) derived from the literature; the membership of the star; and the literature source.  The membership column lists two values: member/non-member (M or NM) from comparing $\gamma$ to the cluster \vhelio\ and yes/no (Y/N) referring to whether the star made the RV cut for the cluster as described in Section \ref{sec:vhelio}. In only one case (NGC~2477~6062) did the combined motion of the binary make a true non-member appear to be a member  from the single epoch RV measured here.  On the other hand, 9 of the 19 true members were flagged as suspected non-members here  because of the added binary motion.

With only a single epoch in radial velocity, the best way to look for binary companions among the suspected members is  to inspect the cross-correlation functions for signs of asymmetry or multiple peaks.  However, this technique can only detect companions bright enough to contribute significantly to the combined flux of the system.  For S/N of 10, the fainter star must contribute more than 10\% of the combined flux to be detectable over the noise, which corresponds to stars up to 2.5 magnitudes fainter than the primary.
  For  red giant primaries, the main sequence stars at  the same color are too faint to contribute to the total light at this level, which means that any companions that are bright enough to contribute significantly  to the total light are also bluer than the RGs. 
The $V$ magnitude  of composite systems  can be as much as 0.75 magnitudes brighter than the isochrone (for equal brightness components) and the colors can be shifted significantly towards the blue by as much as $\sim0.5$ magnitudes if the companion is at the MS turn-off.

Stars with such relatively bright companions should either have colors or magnitudes that are very dissimilar from the isochrones or show asymmetries in the CCF peaks due to the secondaries' contributions to the total light.
 None of the known binaries in Table \ref{tab:oc_binary} have any obvious indication of their companion in the shape of the CCF.  However,  one of the known binaries  appears to have  a   luminosity that is inconsistent with the isochrone. This star is M67~136, which  is a known binary system with a long period ($P\sim1495$ d), discovered by \citet[][S1072 in that study]{1990AJ....100.1859M}. This  example likely has  a subgiant primary, but the authors note that its photometric colors are very difficult to explain even considering the photometric contribution of the companion.    

For the enhanced rotators, there are three that appear ``over luminous'' for their color, implying that a suspected stellar companion is bright enough to appear in the CCFs as a secondary peak or asymmetry in the primary peak. However, after inspecting 
the CCFs of all of the enhanced rotators, we  found that  only NGC6005~4 showed obvious asymmetry. Unfortunately for both clusters with over-luminous fast rotators, the scatter in the photometry indicates the presence of differential reddening, which will also move stars  to brighter magnitudes and  bluer colors if they are less reddened than average. However, a more likely explanation for the lack of a secondary  CCF peak is that the companion's orbital RV is not significantly different from that of the primary star, which is more likely to occur for more distant separations or lower inclination angles.  With the data at hand, we can only say that one enhanced rotator (NGC 6005~4) shows strong evidence of binary companion in both its photometry and CCF asymmetry, whereas the lack of strong evidence for the remaining stars does not necessarily rule out stellar companions.  
Additional radial velocity monitoring is needed to confirm the presence or absence of  a stellar companion and whether it is close enough to have affected the rotation of the primary star.

\subsection{Cluster Rotation Compared to the Field}
Field giant stars also have low fractions of rapid rotators. In the \cite{1999A&AS..139..433D} sample of giants,  2.5\%  (11 of 432)  of the apparently single RGs have intermediate \vsini\ (4--8~\kms, the moderate and modest rotators) while 1.3\% (6 of 432) have \vsini$\geq 8$~\kms. In this study, intermediate and rapid rotators (excluding likely subgiant stars) comprise 4\% and $<0.4$\% of the sample, respectively. Thus, there is no statistical difference between the fraction of rapid and intermediate rotators between the open clusters studied here  and  the \cite{1999A&AS..139..433D} field giant populations. 
\cite{2011ApJ...732...39C} used 10~\kms\ as a cut off for rapid rotation and found 2.2\% of field giants to be rotating this rapidly. 
The higher fraction in \cite{2011ApJ...732...39C}  is likely due to binaries because they did not have multi-epoch RVs to flag binaries, whereas in this study, radial velocity mismatch with the cluster probably excluded some binaries in this work. 
Although the number of rapid rotators is smaller than one would expect from the field giant studies mentioned, the difference is not statistically significant.

\begin{figure*}
\centering
\includegraphics[width=0.49\textwidth]{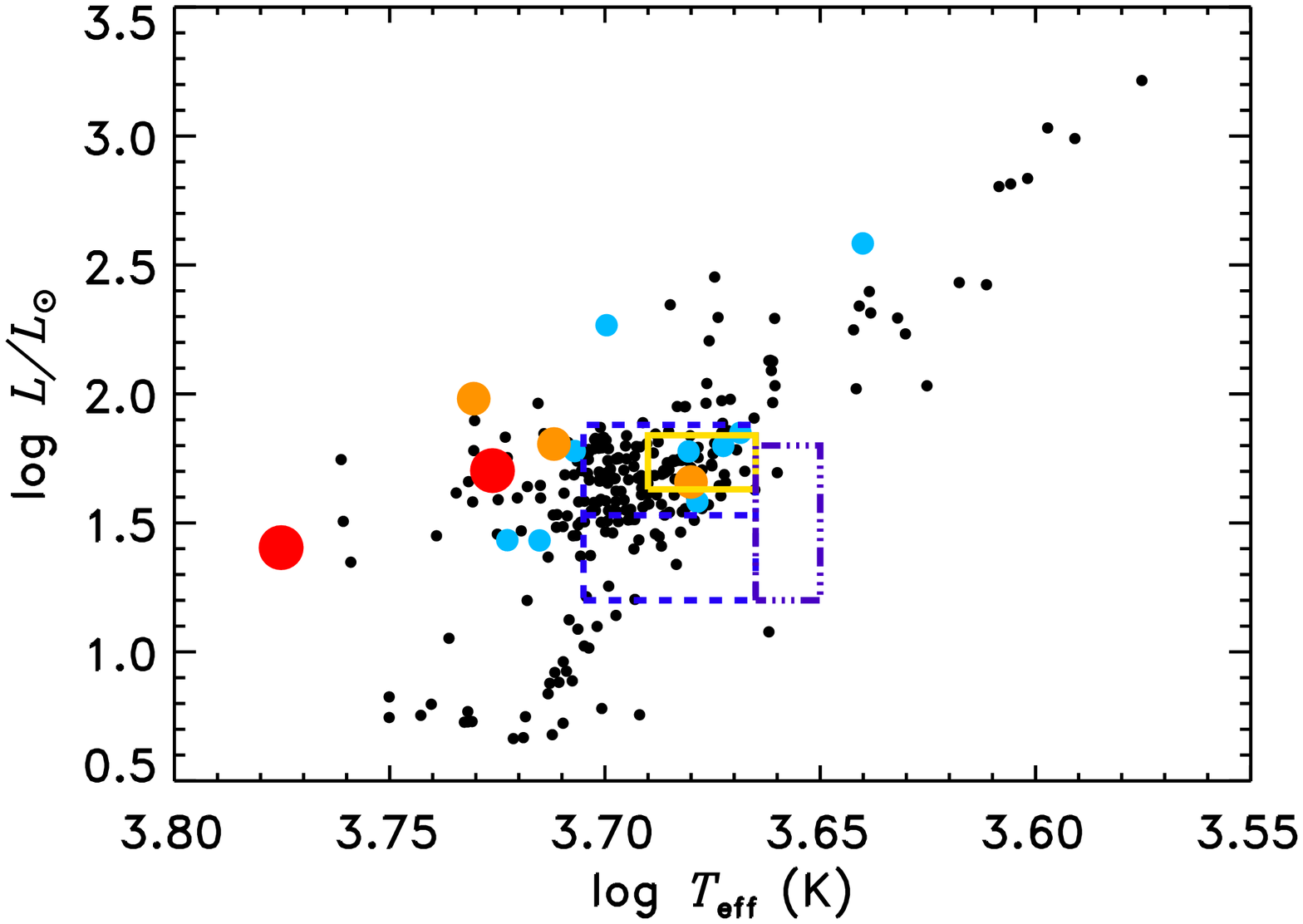}
\includegraphics[width=0.49\textwidth]{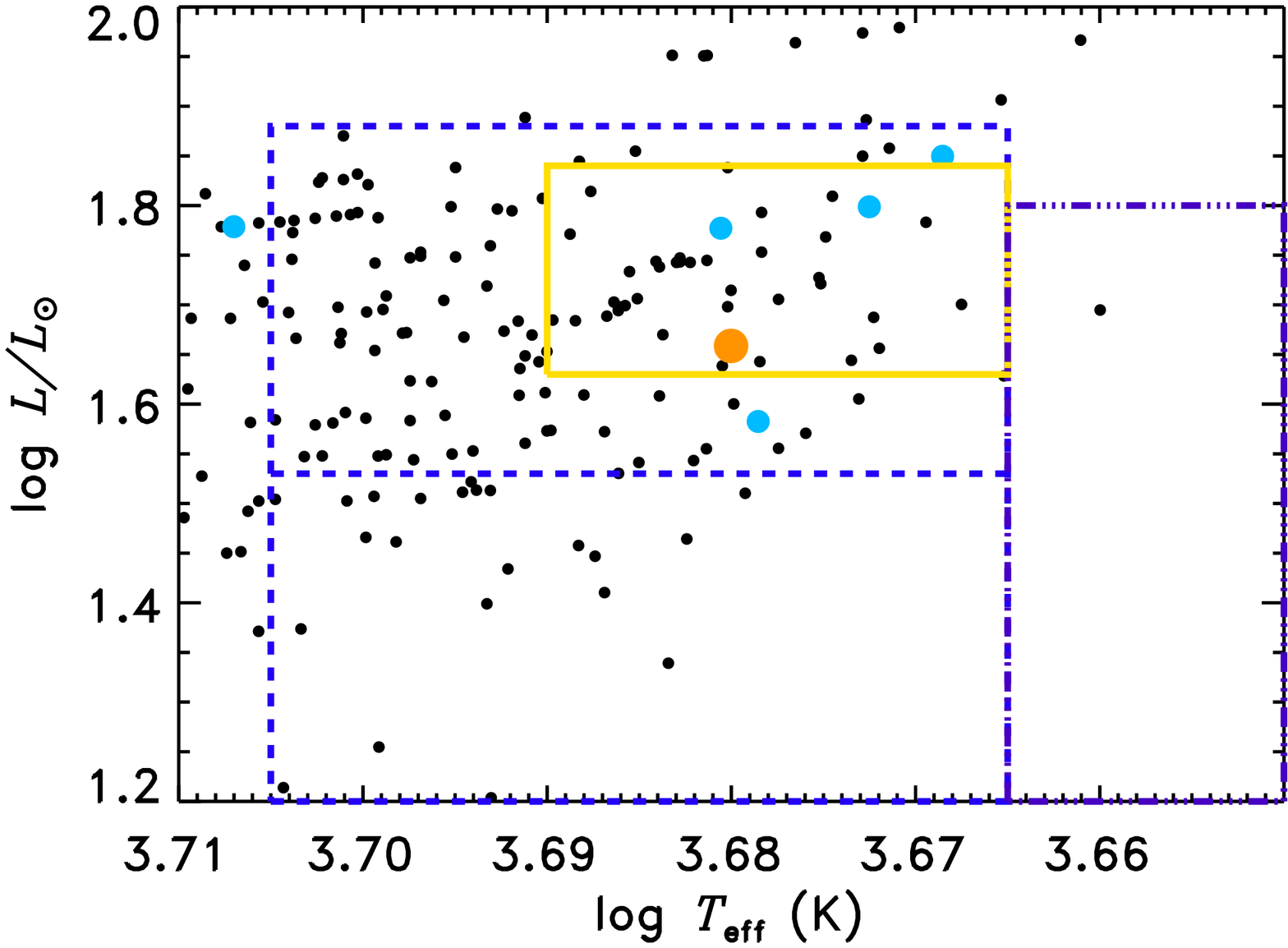}
\caption{\label{fig:vsini_HR}  Hertzsprung-Russell diagram of all of the cluster stars, with the colors and sizes of the points  indicating rotation as in Figure~\ref{fig:iso_rot}. Colors available for each cluster (either $B-V$ or \vmic) have been converted to photometric temperatures 
using the \cite{2000AJ....119.1448H} color-temperature relations for giants, after correcting for reddening.  The left panel shows the full parameter space spanned by the sample, while the right panel shows a blow-up.  The boxes show the same subsets defined in \cite{2008AJ....135..209M} for ease of comparison. (A color version of this figure is available in the online journal.)}
\end{figure*}

 \citet[][hereafter M08]{2008AJ....135..209M} discovered  in their large sample of field giants that enhanced rotators are preferentially found both in the red clump   
and among stars that had just recently completed first dredge-up.   We test for a similar result with our data. Figure~\ref{fig:vsini_HR} shows the distribution of the cluster sample on the HR diagram with boxes showing the sample regions defined in M08. 
The dereddened colors were transformed to stellar effective temperature using the \cite{2000AJ....119.1448H}  relationships for giants. 
The 2MASS colors were first transformed to the CTIO $J-K_S$ system using the updated \cite{Carpenter:2001fv} transformations\footnote{http://www.ipac.caltech.edu/2mass/releases/allsky/doc/sec6\_4b.html}. Luminosities were derived using \cite{Torres:2010gd} $V$-band bolometric corrections.
The large boxes created by the dashed lines delineate the red clump (top box) and a control region just below it (bottom box). 
M08 found enhanced rotation in the top box compared to the lower box, which they argued was due to AM dredge-up. Stars in the lower box are primarily first ascent stars while the top box have a higher fraction of red clump stars. 
The dash-dot  line on the very right edge delineates stars that have just completed first dredge-up.  In M08, the stars in this box also had higher rotation.

 Seemingly consistent with these results, five of the enhanced  rotators fall in the red clump box compared to zero in the comparison box below the red clump box. However, the comparison box is poorly populated and contains only a fifth of the number of stars as in the upper box. Hence, an absence of enhanced rotators in the lower box is still consistent with an equal fraction of enhanced rotators.   
Similarly, the box that samples the completion of dredge-up has a single star, so we cannot test the rotation distribution.  
The reason for the poor sampling in the cluster sample may be due to the different stellar mass distributions. A large fraction of the M08 sample probed masses below $\sim$1.4~\msun.

The solid-line box inlaid in the red clump region defines the portion of the red clump where M08 found two moderate rotators that they argued must have engulfed planets.  This parameter space also has the greatest ambiguity in discriminating  red clump stars from first-ascent giants.  It is of note that the three enhanced rotators in this box are all Pismis~18 stars, 
which suggests that stars of this particular mass might all be experiencing some phase of enhanced rotation that other stars do not, e.g.,  AM dredge-up, which is discussed in Section \ref{sec:AMDU}.

In addition to the enhanced rotators found in the ambiguous  stellar evolution region, there is another group of enhanced rotators that appear to make up the bluest extent of the red clump ($\log T_{\rm eff} \sim 3.72$ and $\log L/L_\sun \ge 1.2$). This parameter space could also be consistent with the post-MS/pre-RGB stage of stars more massive than 2.2\msun; however, all of our clusters are old enough to exclude this mass range.    Thus, it appears that most of the enhanced rotators are quite evolved, despite the fact that  they should have all arrived at the base of the RGB as slow rotators.
The subgiant stars in our sample are represented by two groups  near $\log T_{\rm eff} \gtrsim 3.74$. The subgiants in the younger clusters (at $\log L/L_\sun \gtrsim1.2$) are few in number but have at least one  rapid rotator, while the stars from older clusters (at $\log L/L_\sun \sim0.8$) are generally slower rotators. 

\subsection{Rotation Distributions by Stellar Mass}
Stars more massive than 1.3~\msun\  lose little AM on the MS.
However,  as already described in Section \ref{sec:intro}, the specific AM at birth is larger for stars more massive than 1.6~\msun\ than  for stars with masses between 1.3--1.6~\msun\  \citep{1997PASP..109..759W}.
Nevertheless, all of these intermediate mass stars ($M>1.3$~\msun)  do  eventually develop outer convection zones as they evolve off of the main sequence and  consequently experience strong magnetic braking.  Since higher rotation correlates with higher AM loss, the differences between the faster and slower rotators should diminish as the stars evolve. 
Subgiants are generally sparse in studies on rotation especially in the intermediate mass regime, owing to the short time scale of this stage, and there is little data on whether the rotation distributions eventually merge. 
Even in the subgiant-targeted rotation study by \cite{doNascimentoJr:2003gl},  there are almost no subgiants more massive than $\sim$1.3~\msun\  in the transition from 6000--5200~K ($(B-V)_0\sim 0.55$--0.84), which spans at least half of the temperature range of the subgiant branch. Nevertheless,  \cite{doNascimentoJr:2003gl} clearly find more rapid rotation on the high temperature end of the gap and much slower rotation on the cool side of the gap. 
A quick inspection of Figure~\ref{fig:iso_rot} confirms that the cluster stars follow the same pattern. Once stars reach the base of the red giant branch, they are generally all slowly rotating.  
The growth of  the stellar radius between the MS turnoff and the base of the red giant branch is quite modest---only a 20\% increase for a solar metallicity  1.5~\msun\ star \citep[using models from ][]{Bertelli:2008ge}---not nearly enough to explain the one or two order of magnitude change in the surface rotation.

\begin{figure}
\centering
\includegraphics[width=0.5\textwidth]{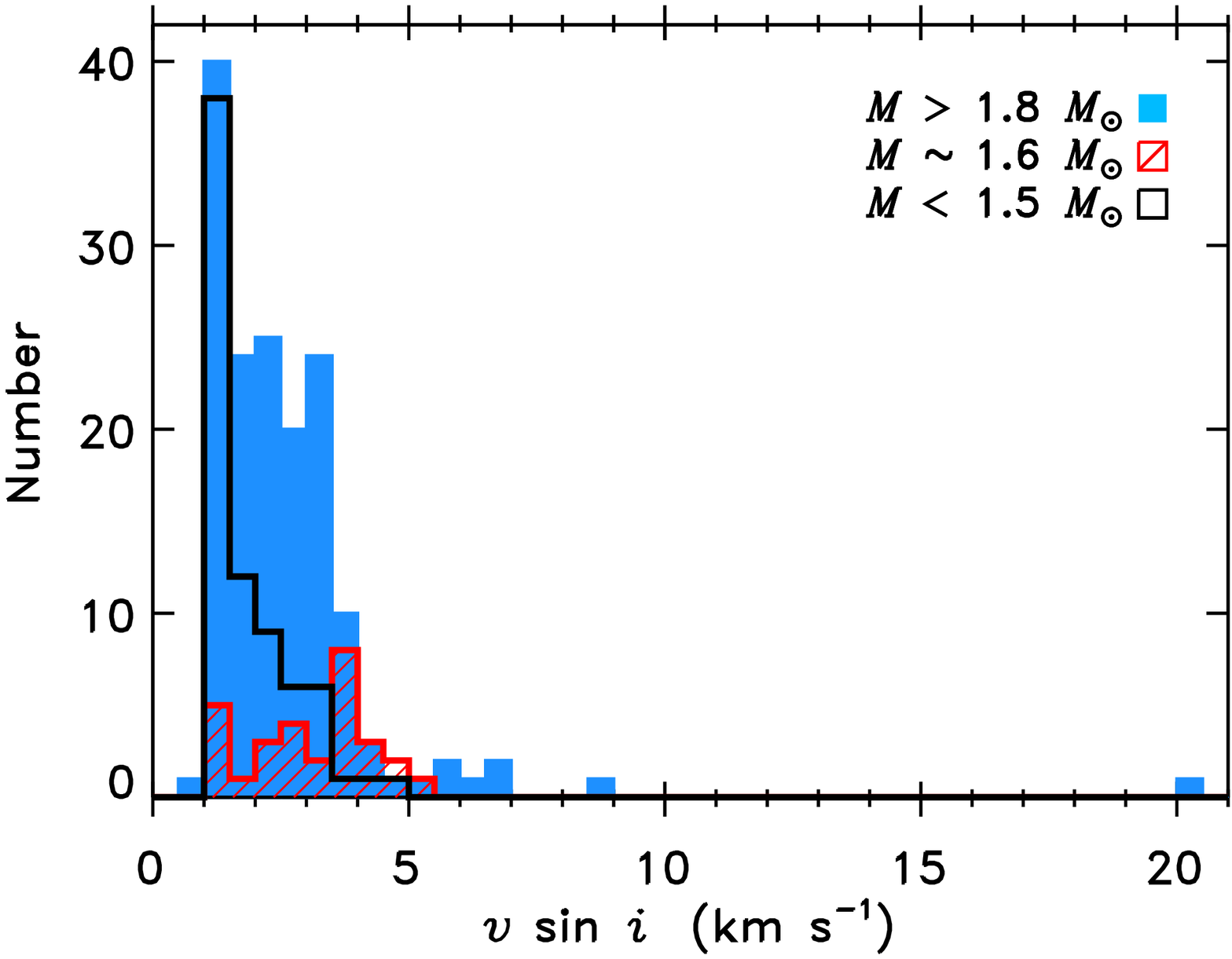}

\includegraphics[width=0.49\textwidth]{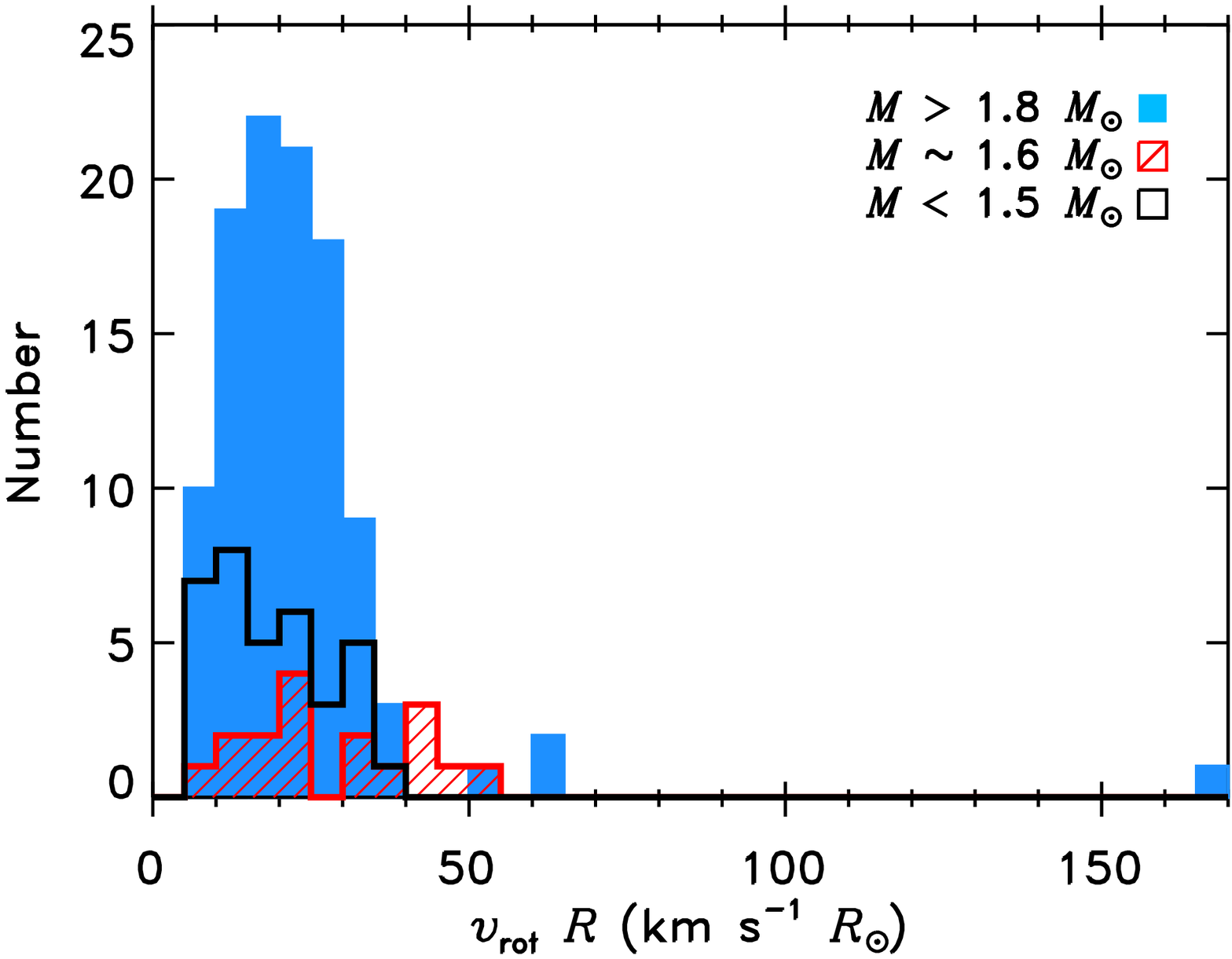}
\caption{\label{fig:vsini_mass} \emph{Top:}  Distribution of \vsini\ for RGs with masses higher (solid histogram) and lower  (open histogram) than 1.6~\msun.  The two clusters with masses right near the transition mass are  plotted separately (hatched histogram). 
\emph{Bottom:}  Distribution of $v_{\rm rot}R$, a proxy for AM, for a restricted sample of red clump stars with $R_\star=7$--11~\rsun.
 (A color version of this figure is available in the online journal.)}
\end{figure}

Our sample can be used  to test whether the initial  differences in observed rotation with mass on the MS are erased by the time
 these stars become RGs or whether these differences persist throughout the RG phase.  Figure~\ref{fig:vsini_mass} shows the distribution of \vsini\ for RGs  above and below 1.6~\msun.   Six clusters are securely above the mass threshold with $M_\star >1.8$~\msun, and three are securely below the threshold having $M_\star < 1.5$~\msun. These separations are true even within the variation of possible ages and metallicities listed in Table \ref{tab:oc_list2}.  The remaining two clusters, NGC~2506 and Ruprecht~147, have mass estimates very close to the transition. The RGs in NGC~2506 in particular may be at, above, or below the transition mass depending on which set of parameters are adopted. The stars in these two clusters are plotted in a separate histogram in Figure~\ref{fig:vsini_mass}.
By comparing these histograms, it is clear  that the more massive stars show a distinct difference in their \vsini\ distribution, having both a higher mean \vsini\ and a broader distribution in \vsini, consistent with the idea that the high mass stars evolved from a population with both a larger and  broader initial  AM distribution. This figure also suggests that the stars with estimated masses near 1.6~\msun\ are likely to be from the higher mass population. 
The two-sided Kolmogorov-Smirnov (K-S) statistic confirms these by-eye deductions.  The secure groups of high and low mass stars have a K-S probability of only  0.004\%. The stars in the transition mass range have a 0.5\% likelihood of being drawn from the high-mass population but only a 0.005\% likelihood of coming from the low mass population.  

Although the \vsini\ distributions are quite different, this  is not necessarily conclusive evidence  that the AM distributions are different.  The   radii of RG stars in this sample range from 2.5 to 94~\rsun, and lower mass stars have much more extended RGBs (i.e., the grow larger) than more massive stars.
 A better set of distributions to compare are those that more directly probe the  AM in the convection zone.
A simple approximation of the specific AM in the envelope is given by  $J_{\rm sp} \equiv J/M \sim I \Omega/M \sim MR^2v_{\rm rot}/RM \sim v_{\rm rot} R$, and the distribution of this quantity is plotted in the second panel of Figure~\ref{fig:vsini_mass}.  
Restricting the sample to  a small range in $R$ will also  mitigate biases introduced at the lowest \vsini.
The histograms in the second panel only contain RC stars with estimated radii of 7--11~\rsun. This selection still  captures 60\% of the entire sample.
  The distributions look much more similar, and a K-S test applied to these two $J_{\rm sp}$ distributions yields a 21\% probability that the distributions are from the same parent population, and the stars near the transition mass have a nearly equal chance of being drawn from the high or low mass star population (5\% in the former and 4\% in the latter case).  Therefore, despite differences in AM imparted at birth for stars in this mass range, there is no conclusive evidence for a difference in the distribution of AM by the time these stars evolve to the red giant stage.

\section{Sources of Excess Angular Momentum}  
\label{sec:sources}
In this section, we briefly discuss possible sources of the enhanced surface AM seen in the fastest rotating stars. 

\subsection{Angular Momentum Dredge-Up}
\label{sec:AMDU}
The onset of surface convection in intermediate mass stars evolving off the MS is thought to be responsible for the rapid deceleration of these stars' surface rotation via magnetic braking. The convection layer deepens through the earliest stages of the red giant branch evolution until as much as 80\% of the stellar mass is convecting.   
\cite{1989ApJ...346..303S} noted that although it was clear that intermediate
mass stars undergo a rapid deceleration in their surface rotation as they crossed the Hertzsprung Gap, it was not clear whether the stars were losing or conserving AM.  If the stellar AM remains nearly constant, it implies that the surface  AM is sequestered and could re-surface during first-dredge up. This scenario is particularly enticing for explaining very rapid rotation (\vsini$>15$~\kms) of intermediate mass RGs because the MS AM of these more massive stars is so high. 
AM dredge-up has also been used in the M08 field giant study discussed previously  to explain a much more modest enhancement of \vsini\ of only a few \kms\ in stars that are mostly less massive than $\sim$1.4~\msun. 
The same argument could also be applied to the enhanced rotators found in the region delimited by the solid-line box in Figure \ref{fig:vsini_HR}.
The fact that these enhanced rotators are found in only one cluster, Pismis 18, and that they share a common evolutionary stage suggests that AM dredge-up may be a common property of 1.9~\msun\ stars---the estimated mass of the RGB stars in Pismis 18.

If the fast rotators in the ambiguous part  of Figure~\ref{fig:vsini_HR} are red clump stars instead of first ascent stars, then any  AM redistribution must be happening much later in the RG's evolution.
Recently, asteroseismic analyses of data from the \emph{Kepler} satellite has resulted in the detection  of the core rotation of RGs and, more importantly, the evolution of that core rotation   \citep{Mosser:2012dj}.  Mixed-mode stellar oscillations couple  the gravity waves in  the RG  core  to the pressure waves in the stellar envelope, and the rotational splitting of these modes  are dominated by the rotation of the stellar core and directly measures  how the core's rotational period changes as the star evolves.  
   \cite{Mosser:2012dj}  found only a small decrease in the rotational splitting as a function of the RGB radius, which can be interpreted as little change in the core's rotational period while the star is on the RBG. However, the red clump stars showed significantly smaller rotational splittings, pointing towards a significant slowing of the core sometime near the end of the RGB stage.  \cite{Mosser:2012dj} speculate that AM is transferred from the core to the outer layers when the star is near the tip of the RBG; however, their sample does not probe the latest stages of RGB evolution. 
   
  These results do provide a means of estimating the surface rotation increase expected in a red clump star from the transport of core AM to the stellar envelope. They find that the 
   average rotation in the core decreases  by a factor of 6.  However,  the ``core region''  is somewhere below 
0.01 $R_\star$ so that the moment of inertia of the core is  $\lesssim 10^{-4}$ the moment of inertia of the envelope, and this implies that  the core rotation must be at least 100 times faster than the envelope to change the envelope's AM by 1\%. In contrast, even the  ``modest'' and ``moderate''  rotation  in the stars studied here corresponds to at least a  doubling to quadrupling of the envelope's AM.  In other words, the observed  level of decreasing core rotation in RGs cannot fully explain the observed level of enhanced rotation in  stellar envelopes. 

\subsection{Externally Supplied Angular Momentum}

If the stellar interior proves to be an unfeasible source of AM, an external origin is needed. 
Synchronous rotation with a close stellar companions is an obvious source, but for  apparently isolated stars, a stellar merger \citep{Hills:1976ug} or substellar engulfment  \citep{1983ApJ...265..972P} scenario is required.
Given the dearth of close orbiting objects with masses intermediate to stars and giant planets, 
the so-called ``brown dwarf desert''  \citep{Grether:2006ek}, mergers  consist of one of two mass regimes.
Stellar mergers will be much more energetic and contribute much more AM than substellar companion engulfment, and such mergers are thought to be the cause of the FK Comae stars, which have ultra-rapid rotation, e.g.  \vsini\ of ~100~\kms\ \citep{1981ApJ...247L.131B}.  Additionally, the mass increase is so large that the now-single star will follow a more massive evolution track. Therefore, the fast rotators labeled as ``over luminous'' in Table \ref{tab:oc_RRs} may represent this formation pathway if they are found to have no close companions. The much slower rotation observed in these stars compared to FK Comae stars implies that the merger happened long ago  and that the stars have spun down to more moderate rotation. 

Planet engulfment will contribute far less AM to the RG than a stellar merger, but the orbital AM of an engulfed planet is still sufficiently large to measurably increase the spin of the stellar envelope. 
\cite{2009ApJ...700..832C} studied the evolution of known planet-hosting systems and found that rapid rotation from the engulfed companions could typically persist for a third of the RG star's evolution. 
However, planets engulfed  late in the RGB evolution would not usually create rapid rotation because  the star's  moment of inertia grows faster than the orbital AM available in the orbits of the planets 
 that are increasingly  ``engulfable'' by tidal orbital decay.  
If, however, the AM gained late in the RGB evolution is preserved as the star re-contracts  in its evolution to the red clump, the AM gained from the planet would be sufficiently large to cause enhanced rotation.     
 \cite{Massarotti:2008id} argued that this was the explanation for the  slightly enhanced rotation seen in the red clump stars originally presented in the M08 sample. 
 If the Pismis 18 stars with enhanced rotation are in fact red clump stars, this engulfment scenario could also be a reasonable explanation provided that the cluster has an unusually large fraction of stars with sufficiently close, large substellar companions.

\section{Summary}
\label{sec:summary}
We surveyed the rotation distribution of RGs in eleven open clusters to search for rapid rotators in an effort to better understand the outliers of this otherwise generally  slow-rotating class of stars.
Rapid and moderately fast rotators appear to be as common  in clusters as in the field population, but the frequency of moderate rotators  in Pismis~18 is surprisingly large.  Separating the stars into mass bins with different average rotation on the main sequence, we find that the more massive stars ($M\gtrsim1.6$~\msun)  have a larger population of enhanced rotators, and the rotation distribution is much broader.  However, when converting to AM and restricting the comparison to stars with similar sizes, the differences, while still present, are not statistically significant. 
Nevertheless, there are clear outliers to the general AM distributions among the higher mass stars. 
 
These data  provide constraints on  models of stellar rotation of intermediate mass stars, which appear to spin down very rapidly during the subgiant evolution phase. Stars in the subgiant phase are quite rare, and their radii do not change drastically enough to account for the  amount of spin-down necessary to explain the slow rotation seen in the earliest RG phase. If some of the AM is sequestered 
in the stars instead of lost by magnetized winds, evolution models can use the rotation distribution of RGs to test how that AM may resurface. 
Measuring additional rotational velocities or rotation periods of intermediate mass stars crossing the Hertzsprung Gap remains a difficult but important piece of the puzzle.
 Distinguishing between internal and external sources of enhanced surface rotation may come from disentangling the ambiguous stage of evolution that may either be first dredge-up or the red clump. AM dredge-up may be at work in the former case, while the latter may require an external AM origin.  Testing these two possibilities could be accomplished with 
 asteroseismic analysis of a rapidly rotating RG to distinguish whether the He core is inert or is an active region of nucleosynthesis. 

\acknowledgments
The author would like to thank Alycia Weinberger for  providing constructive comments on an earlier draft of the manuscript and
the anonymous referee for providing valuable critiques that improved the presentation of the results of this paper. 
This research has made use of the WEBDA database, operated at the Institute for Astronomy of the University of Vienna and  the VizieR catalogue access tool, CDS, Strasbourg, France. The original description of the VizieR service was published in A\&AS 143, 23.

\end{document}